\documentclass[preprint,3p]{elsarticle}
\usepackage{hyperref} 
\pdfoutput=1

\usepackage{graphicx}
\usepackage{dcolumn}
\usepackage{bm}
\usepackage[mathlines]{lineno}
\usepackage{color}
\usepackage[caption=false]{subfig}
\usepackage{multirow}
\usepackage{amsmath}
\usepackage{algorithmic} 
\usepackage{algorithm}
\usepackage{paralist}

\newtheorem{theorem}{Theorem}[section]

\newtheorem{definition}[theorem]{Definition}

\graphicspath{{figs/}}
\DeclareGraphicsExtensions{.pdf}

\journal{Safety Science Journal}

\begin{document}

\begin{frontmatter}

\title{Modelling dynamic route choice of pedestrians to assess the criticality of building evacuation}

\author[FZJ]{Armel Ulrich Kemloh Wagoum \corref{cor1}}
\ead{u.kemloh@fz-juelich.de}

\author[FZJ,WUPPERTAL]{Armin Seyfried}%
\ead{a.seyfried@fz-juelich.de }

\author[FZJ,WUPPERTAL]{Stefan Holl}%
\ead{st.holl@fz-juelich.de }

\address[FZJ]{J\"ulich Supercomputing Centre, Forschungszentrum J\"ulich GmbH, 52428 J\"ulich, Germany}
\address[WUPPERTAL]{Computersimulation f\"ur Brandschutz und Fu\ss g\"angerverkehr, Bergische Universit\"at Wuppertal, Pauluskirche 7, 42285 Wuppertal, Germany}

\cortext[cor1]{Corresponding author, Tel:+49-2461-614193, Fax:+49-2461-616656}

\begin{abstract}
This paper presents an event-driven way finding algorithm for pedestrians in an evacuation scenario, 
which operates on a graph-based structure. The motivation of each pedestrian is to leave the facility. 
The events used to redirect pedestrians include the identification of a jam situation and/or identification of a better route than the current.
 This study considers two types of pedestrians: familiar and unfamiliar with the facility. 
Four strategies are modelled to cover those groups. The modelled strategies are the shortest path (local and global); 
They are combined with a quickest path approach, which is based on an observation principle. In the quickest path approach, 
pedestrians take their decisions based on the observed environment and are routed dynamically in the network 
using an appropriate cost benefit analysis function. The dynamic modelling of route choice with different 
strategies and types of pedestrians considers the manifold of influences which appears in the real system and raises questions 
about the criticality of an evacuation process. To address this question criteria are elaborated. 
The criteria we focus on in this contribution  are the evacuation time, the individual times spent in jam, 
the jam size evolution and the overall jam size itself. The influences of the different strategies on those evaluation 
criteria are investigated. The sensibility of the system to disturbances (e.g. broken escape route) 
is also analysed. 
\end{abstract}


\end{frontmatter}


\section{\label{sec:introduction}Introduction}
In recent years pedestrian dynamics has gained more importance and a lot of attention due to continuously growing urban population and cities 
combined with an increase of mass events. This sets new challenges to architects, urban planners and organizers of mass events. One of the main goals is an effective use of the designed facility, 
for instance by minimizing jams thereby optimizing the traffic flow. In this context pedestrian simulations are already used, e.g. for escape route design 
\cite{Exodus,TraffGo2005,AseriHand,Thompson1995}. However the approaches used in these simulations are only the first step to model the manifold influences on human 
beings during an evacuation process. Major issues in this area include orientation and way finding: Given a set of possible routes, 
which criteria influence pedestrians choice for a particular route? This is essential for reproducing route choice in computer models and is 
difficult due to the many underlying subjective influences on this choice. The manner by which pedestrians choose their way has a direct 
influence not only on the overall evacuation time but also on the average time pedestrians spend in a jam. 
In this paper we restrict ourselves to the case where pedestrians all have the same motivation, to leave the facility. 

The approaches of modelling pedestrians motion fall into two main groups: microscopic and macroscopic models. Microscopic models are further 
categorized in spatially discrete (e.g. Cellular Automata \cite{Kirchner2003a,Blue2001,Nishinari2006}) and spatially continuous models 
(e.g. Social Force Model \cite{Helbing1995}, Generalized Centrifugal Force Model \cite{Chraibi2010a}). For a detailed overview we refer 
to \cite{Schadschneider2009a}. CA models use floor fields (static and/or dynamic) to direct pedestrians to a destination point. Route choice in continuous models can be achieved by means of a network which consists of a set of destination 
points. This type of way finding is known as graph-based routing 
\cite{Geraerts07, Barraquand96, Dijkstra2002, Molnar1995, Lovas94}. The destination points can be pre-determined (exits for instance) or adjustable (crossings, turning point at the end of a corridor for instance). The minimal network is usually a 
a visibility graph (see \cite{BergCheong2008} for more details)
which ensures that any location on the facility is within the visibility range of at least one node. The initial network can be 
refined by adding more adjustable points converging to CA \cite{Asano2010}. This graph can be extended to a navigation graph by
adding more adjustable points. The generation of such graph is a complex process, some efficient methods are presented in \cite{Hoecker2010,Alt1988}. Once built the shortest path is usually determined using well established algorithms such as Dijkstra \cite{Dijkstra1959} or Floyd warshall \cite{floyd1962}. Such networks are widely spread in motion planning by robots as well.

The intrinsic behaviour of humans in the case of an evacuation is generally to 
follow the seemingly (self estimated) quickest path. 
This is indeed a subjective notion as it depends on some prerequisites, e.g. whether or not the pedestrian is familiar with the facility. 
The modelling of the quickest path is achieved by systematically avoiding congestions. In CA this is achieved by means of dynamic
floor fields \cite{Kirchner2002} where pedestrians moving increase the probability of using that path thereby making it more attractive for 
other pedestrians. This implicitly leads to congestion avoidance. The density in front of the moving pedestrians within their sight range 
is also considered \cite{Kirik2009} as well as the payload at exits \cite{Zhao2010}; other approaches include navigation
fields \cite{Kretz2009,Hartmann2010}. Continuous models usually optimize the travel time in the constructed network. In \cite{Hoogendoorn2002}, 
pedestrians minimize their travel time by solving the Hamilton-Jacobi-Bellman equation yielding to the optimal pedestrians path at each time step. A combination
of a graph-based routing with CA is presented in \cite{Hoecker2010} where the fastest path for pedestrians is computed using a heuristic $A*$ algorithm \cite{Russell2003}.

In this contribution an event driven way finding in a graph-based structure is  introduced.
The  approach is based on an observation principle, pedestrians observe their environment 
and take their final decision based on the obtained data. With this observation the quickest path
is achieved. The modelled strategies are the local shortest path (LSP), the global shortest path (GSP), the local shortest combined with the quickest path (LSQ) and finally the global shortest combined with the quickest path (GSQ).

An important point less discussed is given by the criticality of an evacuation process, first and foremost the
meaning of criticality for an evacuation process. Evaluation criteria like the building itself, 
the population size and the initial distribution of the evacuees, the evacuation time are discussed in \cite{Lovas98b}.
More individual criteria like the individual travel time and waiting time are investigated in \cite{Fang2011}.
The most used criteria are the overall evacuation time and a visualisation of the evacuation process at specific times. 
In this paper we elaborate other criteria to address the criticality of an 
evacuation simulation. The analysed criteria are the individual time spent in jam, the jam size evolution over time and the 
total jam size defined as the area under the jam size evolution. 
We give a special credit to the time pedestrians spend in jam as well as the jam size. This is of particular importance. 
In the case of the Hermes project \cite{Seyfried}, there are congestions that vary (in place and size) depending on the 
type of people attending the events. We try to reproduce this observed phenomenon for the forecast of the evacuation dynamic.
 In order to achieve this, route choice has to be individually modelled. Those events involve pedestrians familiar and unfamiliar
with the facility. The four previously mentioned strategies are proposed to reproduce their route choice.  Their influences on the previous mentioned criteria are investigated.

The modelling approaches are presented in the second section.
The third section describes the evacuation assessment criteria.
The analysis of the results including simulation, distribution of the evacuation 
time and distribution of the individual times spent in jam 
using different initial conditions are discussed in the fourth section.

\section{\label{sec:modelling}Modelling}

The framework used for describing pedestrian traffic can be divided in a three-tiers structure. One distinguishes between the strategic, 
the tactical and the operational level \cite{Hoogendoorn2002}. In our model the operational level of the pedestrian walking is described by 
the Generalized Centrifugal Force Model \cite{Chraibi2010a} which operates in continuous space. In this model pedestrians are described by ellipses. The semi-axes of the ellipses are velocity dependent, faster ellipses need more space
in the moving direction. The fundamental diagram is reproduced by the 
model at corridors making it adequate for the analysis presented here. In this paper we focus on the strategic level only, i.e. 
the pedestrians are solely given the next destination point, which is the next intermediate destination for the self estimated optimal route. This is also the direction of the driven force. 
We simulate pedestrians that are familiar with the facility and pedestrians
 unfamiliar with the facility. For those two groups we identify criteria to reproduce their route choice. Those criteria are the local/global 
shortest path and the quickest path. The dynamic change of the strategies is modelled. This emulates the internal state of the pedestrians and 
the strategies are subject to change during the evacuation process. In addition one of the challenges consists of finding a good balance between 
the number of parameters and the numbers of criteria. The model should be as simple as possible with the parameter space kept as small as possible 
while considering as many criteria as required. This is important for model understanding and stability.

\begin{figure}
\centering
\includegraphics[width=80mm]{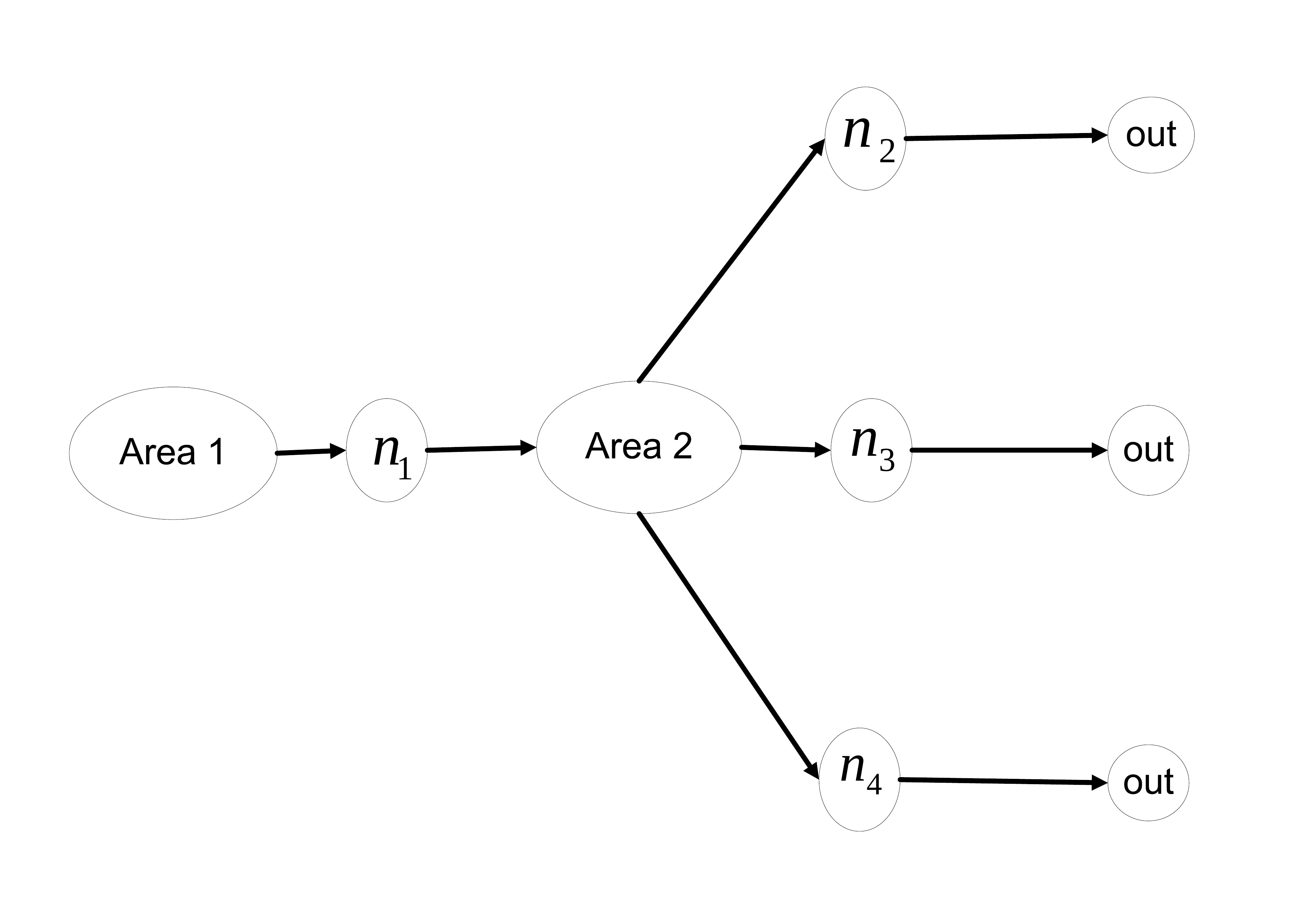}
\caption{Network mapping of the facility presented in Fig. \ref{fig:reference_selection} with 2 decisions areas and 4 nodes.}
\label{fig:decision_area}
\end{figure}

In the framework used here, pedestrians move from one decision area to the next one. The decision areas are connected with nodes, which will be
interchangeable with destination points.
 A decision area is a place where the pedestrian decides which way to go or change their current destination. In this work a decision area is an abstraction
for rooms and corridors, in addition we  restrict ourselves to the case where the destination points are exits and corridors end.
Fig. \ref{fig:decision_area} illustrates this principle and is a mapping 
of the facility presented in Fig. \ref{fig:reference_selection}. The pedestrians will be moving from the decision area 1 to the decision area 2. 
The two areas are connected with one node $n_1$. The network is automatically generated from the facility based on the inter-visibility of the exits.
The Euclidean distance between the nodes is used as weights.

\subsection{Shortest Path}
The most straight-forward routing approach is the local shortest path. Once a node in the network is reached, 
the local nearest node is chosen 
as next destination. The global shortest path is determined by running the Floyd Warshall algorithm with path reconstruction \cite{floyd1962} 
on the built graph network. The runtime of $\mathcal{O}(n^3)$ is not an issue since we only have a small amount of nodes. From every node on the graph, 
the global shortest path to reach the outside can be determined. Pedestrians familiar with the facility have the global map i.e. 
the constructed network and may approximate the global shortest path to their final destination independently of their current location.
 They have a better analysis possibility of the current situation. Other pedestrians without any global information choose the local shortest path.  

\subsection{Quickest Path}
\begin{figure}
\centering
\includegraphics[width=80mm]{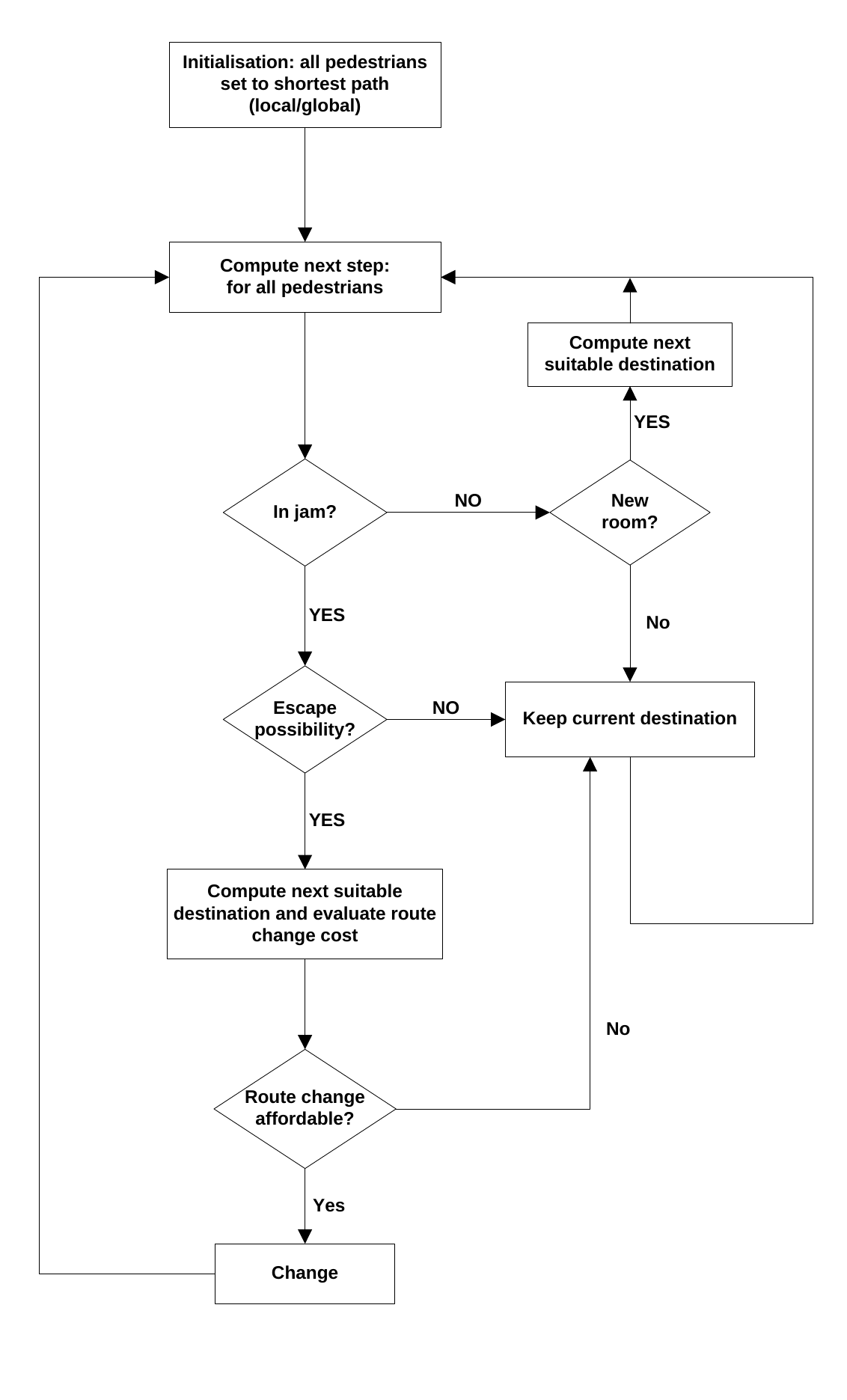}
\caption{Business logic of the quickest path.}
\label{fig:quickest_path}
\end{figure}

In contrast to the shortest path, the quickest path is dynamic and changes with time throughout the simulation. 
The business logic of the routing algorithm is shown in Fig. \ref{fig:quickest_path}. The main events used in this routing 
algorithm to redirect pedestrians are the entering a new room and the identification of a jam situation. 
The pedestrians are first routed using the shortest path, global or local depending on theirs affiliations \cite{KemlohWagoum2010}. 
The key elements of a quickest path routing approach are the estimation of the travel time, 
the estimation of the gain and an assessment of this gain.  Three functions are developped in this paper to model those key elements.
We first define four values which will be used throughout this section.

\begin{definition}{Re-routing time $t_r$}\\
The re-routing time $t_r$ for a pedestrian  $i$ with position $\vec{x}_i$ is the time where one of the following conditions holds:\\
$
\|\vec{v}(t)\| \leq v_{min} ,\forall t \in [t_r-t_{min}, t_r] \: \:
\vee \: \:
\| \vec{x}(t_r) - \vec{n_i}\| \leq d_{min}
\label{eq:events}
$
where $v_{min}$ is the threshold jam velocity, $t_{min}$ the patience time and $d_{min}$  the minimal distance to the node to consider it as reached.
\label{def:event}
\end{definition}

\begin{definition}{Reference pedestrian $Ref(i,\vec{n}_i) $}\\
Let $i$ and $j$ be two pedestrians with positions $\vec{x}_i$ and $\vec{x}_j$.
$j$ is a reference pedestrian to  $i$ with respect to the node $\vec{n}_i$ of the graph and is defined as:\\
$
Ref(i,\vec{n}_i)=\{ j: j \in \mathcal{Q}(\vec{n}_i), \vec{x_j} \text{ visible from } \vec{x}_i \wedge \|\vec{x_j}-\vec{x}_i\|=min \}
$, where $\mathcal{Q}(\vec{n}_i) $ is the jamming queue at the node $\vec{n}_i$. If more than one pedestrian satisfy the condition, one is randomly chosen.
\label{def:reference}
\end{definition}

\begin{definition}{Jamming queue $\mathcal{Q}(\vec{n}_i)$}\\
Let $\vec{n}_i$ be a node in the navigation graph, 
the jamming queue $\mathcal{Q}(\vec{n}_i)$ at the node $\vec{n}_i$ is defined as\\
$
\mathcal{Q}(\vec{n}_i)=\{\forall i: \vec{n}_i \text{ destination of the pedestrian } i \wedge \|\vec{v}_i\|\leq v_{min} \}
$ where $v_{min}$ is the threshold jam velocity.
\label{def:queue}
\end{definition}

\begin{definition}{Visibility range $\mathcal{V}(\vec{x}_i)$}\\
$
\mathcal{V}(\vec{x}_i)$ represents the set of all nodes   
 within the visibility range of the pedestrian $i$  considering the actual location $\vec{x}_i$ in the facility and other pedestrians. 
It is determined using the Algorithm ~\ref{alg:visibility}.
\label{def:visibility}
\end{definition}

\begin{figure}
\centering
\includegraphics[width=80mm]{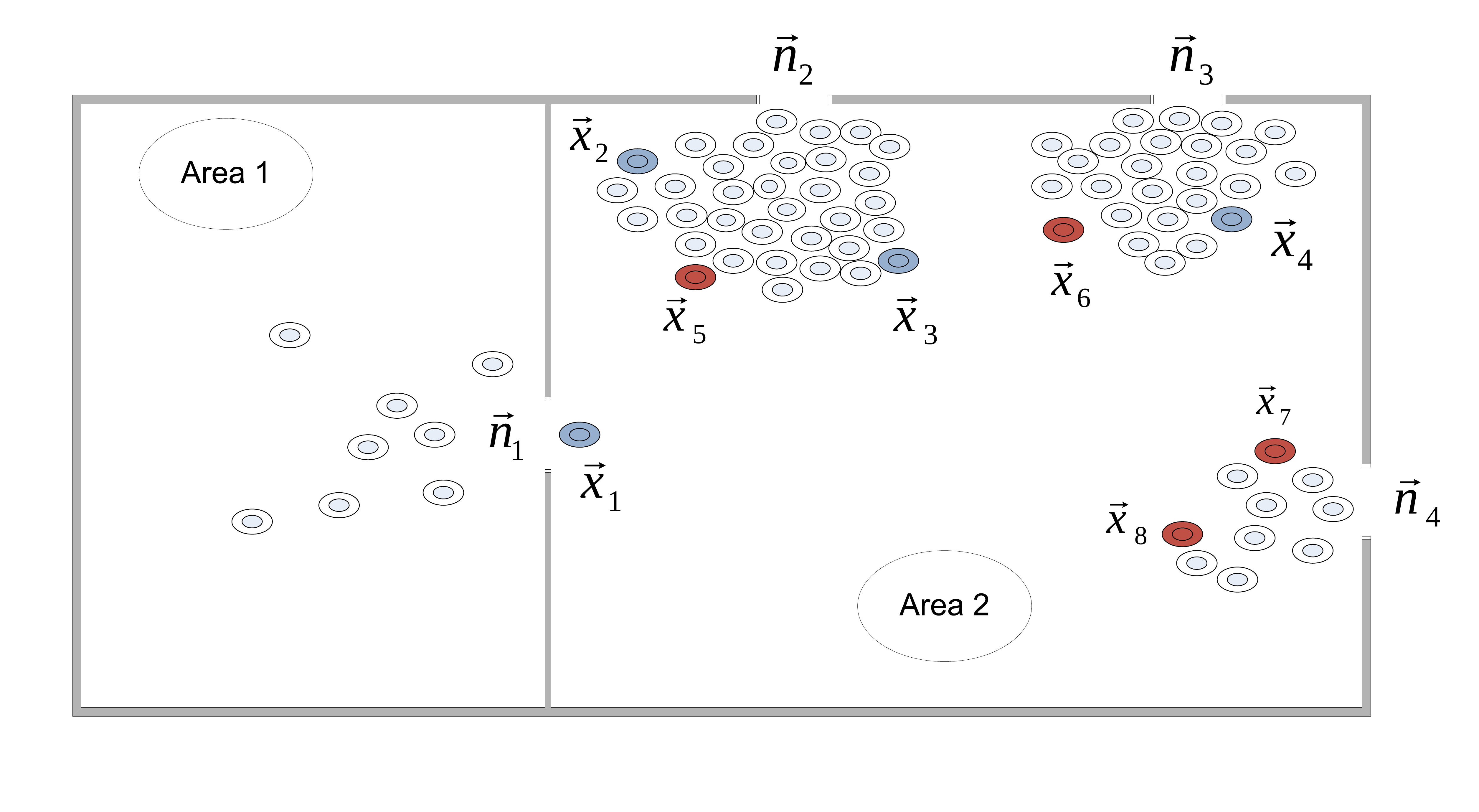}
\caption{Process of selecting a reference pedestrian prior to a route change. Pedestrians are denoted with their positions. $ \vec{x_1}$ will select $ \vec{x_5}$, $ \vec{x_6}$ and $ \vec{x_8}$. 
$\vec{x_2}$ has no clearance of the  current situation and will not select any. $ \vec{x_3}$ selects $ \vec{x_6}$ and $ \vec{x_8}$. $ \vec{x_4}$ will only select $ \vec{x_7}$. 
}
\label{fig:reference_selection}
\end{figure}

At any time $t_r$ (see Definition in \ref{def:event}) during the simulation, a new orientation process is started for the pedestrian $i$.
A reference pedestrian $j$ (see Definition~\ref{def:reference}) is selected  from the queue (see Definition~\ref{def:queue}) and observed during an observation time $t_{obs}$.
At the end of the observation, the expected travel time $t(\vec{x}_i, \vec{n}_j)$ via all nodes in the visibility range $ \mathcal{V}(\vec{x}_i)$ (see Definition~\ref{def:visibility}) is 
approximated using Eq. \ref{eq:time}.

\begin{equation}
t(\vec{x}_i, \vec{n}_j)=
\begin{cases}
\frac{\| \vec{x}_i -\vec{x}_j\|}{\| \vec{v}_i\|} + 
  \frac{\| \vec{x}_i-\vec{n}_j\|}{\| \vec{v}_{ja}\|}, 
  &\text{if a reference $\vec{x}_j$ with average velocity $\vec{v}_{ja}$ was found;}\\\\
\frac{\| \vec{x}_i-\vec{n}_j\|}{\| \vec{v}_i\|}, &\text{if the node $\vec{n}_j$ is free (there is no reference in that case).}\\
\end{cases}
\label{eq:time}
\end{equation}

\vspace{0.5cm}

\begin{equation}
 \|\vec{v}_{ja}\|= \frac{1}{t_{obs}}\int_{t_r}^{t_r+t_{obs}} v_j(t)dt
\label{eq:observation}
\end{equation}

$\vec{v}_{ja}$ is the average velocity of the reference pedestrian over the observation time  defined by Eq. \ref{eq:observation}.
$t_{obs}$ is randomly chosen between 1 and 3 seconds, the minimal distance $d_{min}$ is set to 0.20 $m$ and the minimal jam velocity $v_{min}$ is 0.2 $ms^{-1}$.
The  estimated travel time is converted to a gain using Eq. \ref{eq:gain}.

\begin{equation}
 g(\vec{x}_i, \vec{n}_j)=\frac{1}{t(\vec{x}_i, \vec{n}_j)}
\label{eq:gain}
\end{equation}

The cost benefit analysis (cba) function defined in Eq. \ref{eq:cba} determines whether it is worth changing the 
current destination. $g_1$ and $g_2$ are the gains calculated with Eq. \ref{eq:gain}. $g_1$ is always the current destination and $g_2$ the other evaluated alternatives. 
The benefit returned should be greater than a threshold ($g_{min}$) in order for the pedestrian to consider the change. 
The thresholds taken here are 0.20 for familiar and 0.15 for unfamiliar pedestrians.  
 
\begin{equation}
 cba(g_1,g_2)=\frac{g_1-g_2}{g_1+g_2}
\label{eq:cba}
\end{equation}
 
When a pedestrian is caught in a jam  for a period $t_{min}$ (patience time) which varies depending on the pedestrian, 
he/she looks for alternatives in the decision area and  in the sight range as described in Fig. \ref{fig:reference_selection}. 
The initial value for  $t_{min}$ is 10 seconds. It is increased by 1 second with any unsuccessful attempt to escape the jam. 
The value is kept until the room is changed. This amortizes the number of routes  changes in the same room.

The process of selecting a reference pedestrian is explained in Fig. \ref{fig:reference_selection}. 
The pedestrian $\vec{x}_1$ has entered a new room. The pedestrians $\vec{x}_2$, $\vec{x}_3$ and $\vec{x}_4$ have identified a jam situation. 
Pedestrian $\vec{x}_1$ selects the reference pedestrians $\vec{x}_5$, $\vec{x}_6$ and $\vec{x}_8$ for the exits $\vec{n}_2$, $\vec{n}_3$ and $\vec{n}_4$ respectively. 
The pedestrian $\vec{x}_4$ has a restricted visibility and will have only $\vec{x}_7$ as reference. There is no possibility 
for $\vec{x}_2$ to change. The references selection is based on the Euclidean distance and visibility range. The visibility is implemented by drawing a line from the concerned pedestrian to the pedestrians in the queue, there should 
not be any intersections with other pedestrians or walls in the room. It is important to mention here, that the queue size does not play 
a major role, more important is the processing speed of the queue.

\begin{algorithm}
\caption{Visibility Range of Pedestrian $\vec{x}_i$: $\mathcal{V}(\vec{x_i})$}
\textbf{Input:}  Pedestrian $\vec{x}_i$\\
\textbf{Output:} $\mathcal{V}(\vec{x_i})$ 
\begin{algorithmic}
\STATE {$\mathcal{P} \gets $ all pedestrians $\setminus$ \{$\vec{x}_i$\}}

\STATE {$\mathcal{W} \gets $ all nodes connected to the actual decision area of $\vec{x}_i$}

\STATE {$\mathcal{V} \gets \emptyset$} 

\STATE {$obstacles\_min \gets 2$} 
\FORALL{nodes $\vec{n}_j$ in $\mathcal{W}$}  
  \STATE {$obstacles \gets 0$}
  \STATE {$ u \gets $ queuing pedestrians at node $\vec{n}_j$}

  \FORALL{pedestrians $\vec{x}_j$ in ${u}$ } 
  \FORALL{pedestrians $\vec{x}_k$ in $\mathcal{P} \setminus \{\vec{x}_j\}$ }  
    \IF {$[\vec{x}_i\vec{x}_j]$ intersects with the ellipse shape of the pedestrian $\vec{x}_k$ }
	\STATE {$obstacles \gets obstacles+1$} 
    \ENDIF
\ENDFOR
 \IF {$obstacles \leq $ obstacles\_min $\AND$ $ \vec{n}_j \notin \mathcal{V} $}
     \STATE {$\mathcal{V} \gets \mathcal{V} \cup \{\vec{n}_j\}$} 
  \ENDIF 
\ENDFOR
\ENDFOR
\end{algorithmic}
\label{alg:visibility}
\end{algorithm}

Unlike other algorithms, the approach presented here is not specialized to a particular case (asymmetric exits choice for instance), 
i.e. not bounded to the geometry. It is also not dependent on the initial distribution of the pedestrians.

\section{Evacuation assessment}
Usually evacuation processes are assessed with a visual proof and evacuation time within a feasible range. One less discussed question 
is the criticality of an evacuation process. The state of an evacuation can be critical for a certain group of the population, 
aged persons for example, but rather harmless for a different group. Also the same results can be interpreted differently 
depending on the surrounding conditions. In this section we address three criteria  to assess an evacuation scenario. 

\begin{figure}[htb]
\centering
\subfloat[Local shortest path]{\label{fig:LSP_DEMO}\includegraphics[width=35mm]{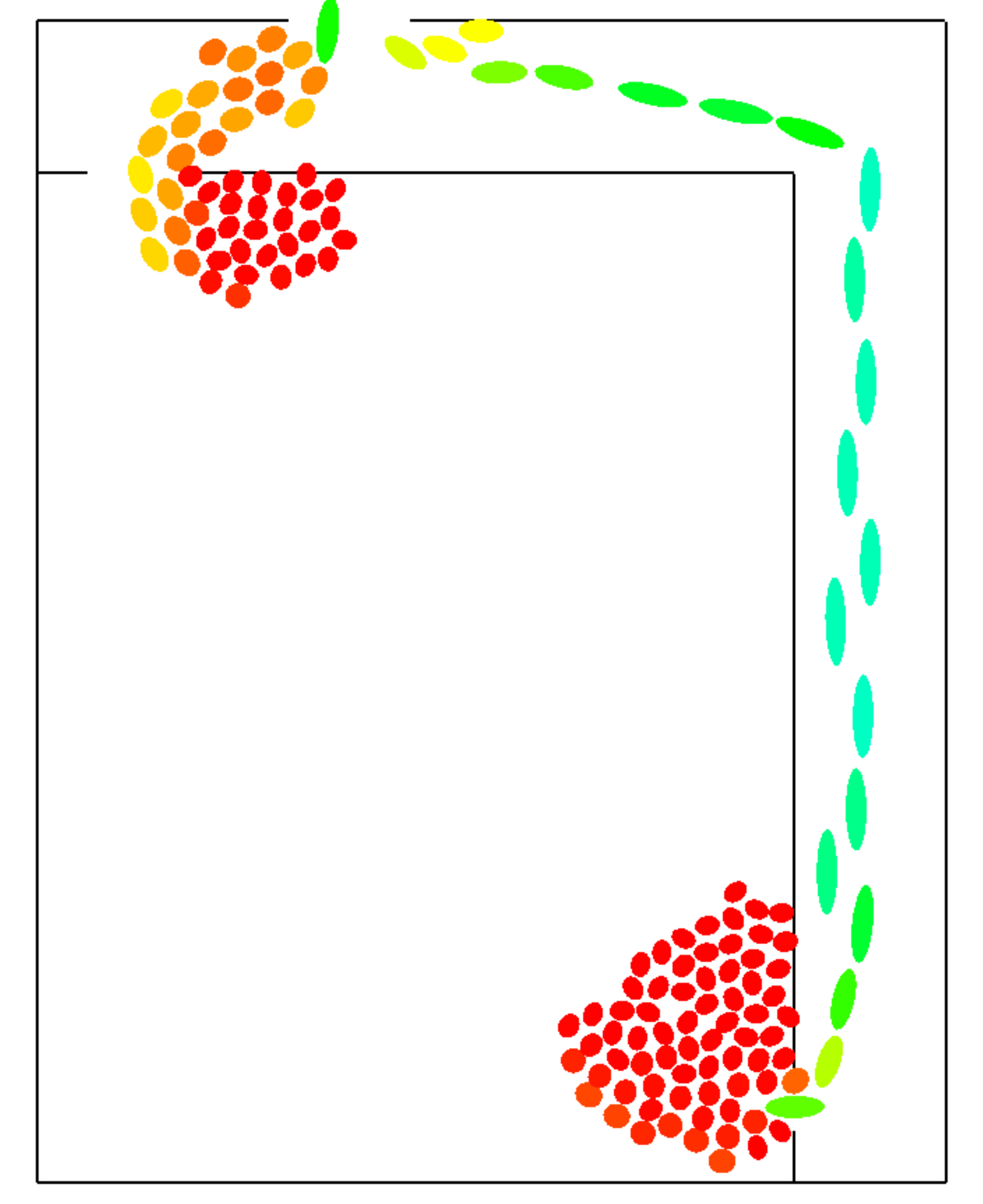}}
\qquad
\subfloat[Local shortest with quickest path]{\label{fig:LSQ_DEMO}\includegraphics[width=35mm]{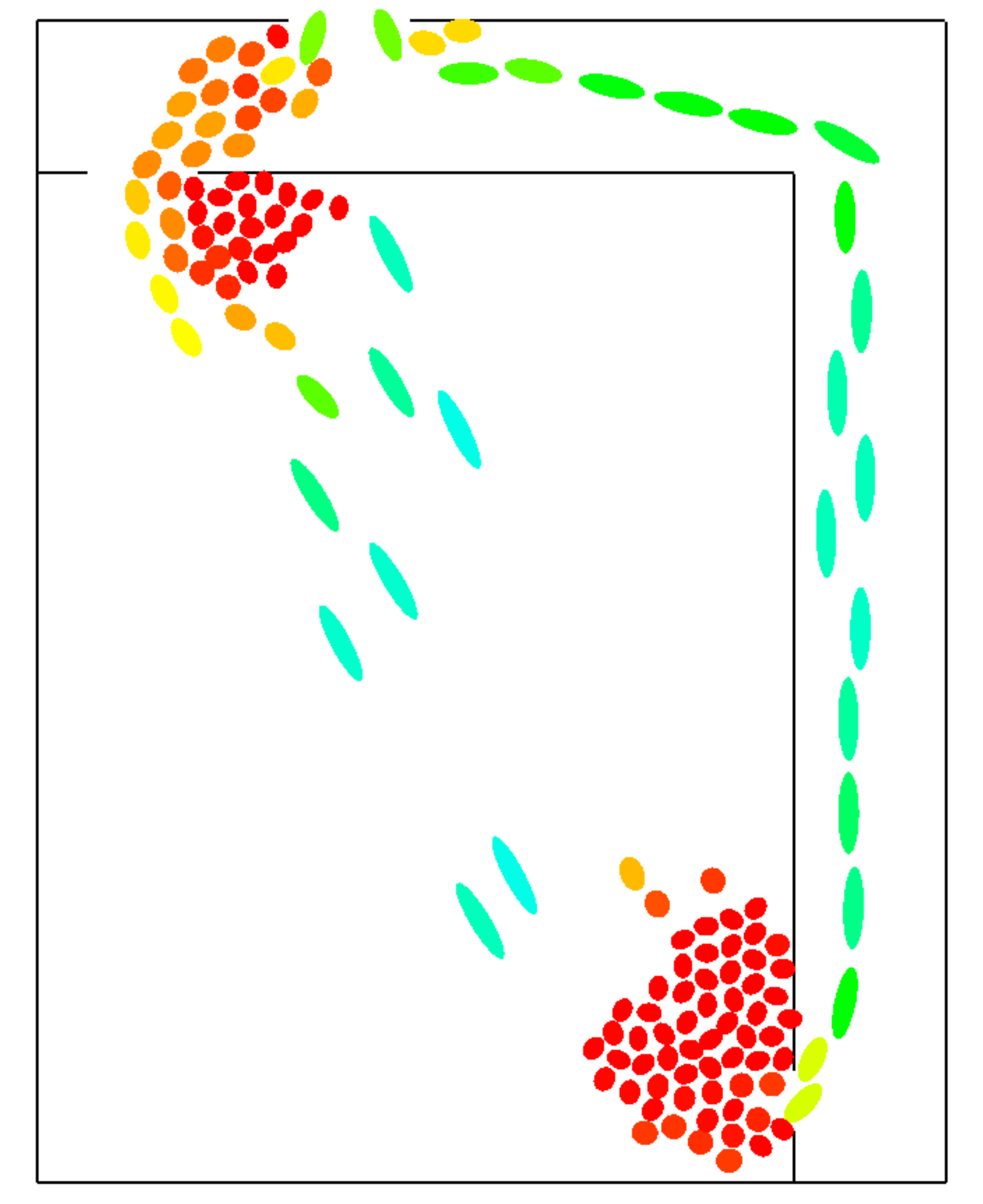}} \\
\subfloat[Global shortest path]{\label{fig:GSP_DEMO}\includegraphics[width=35mm]{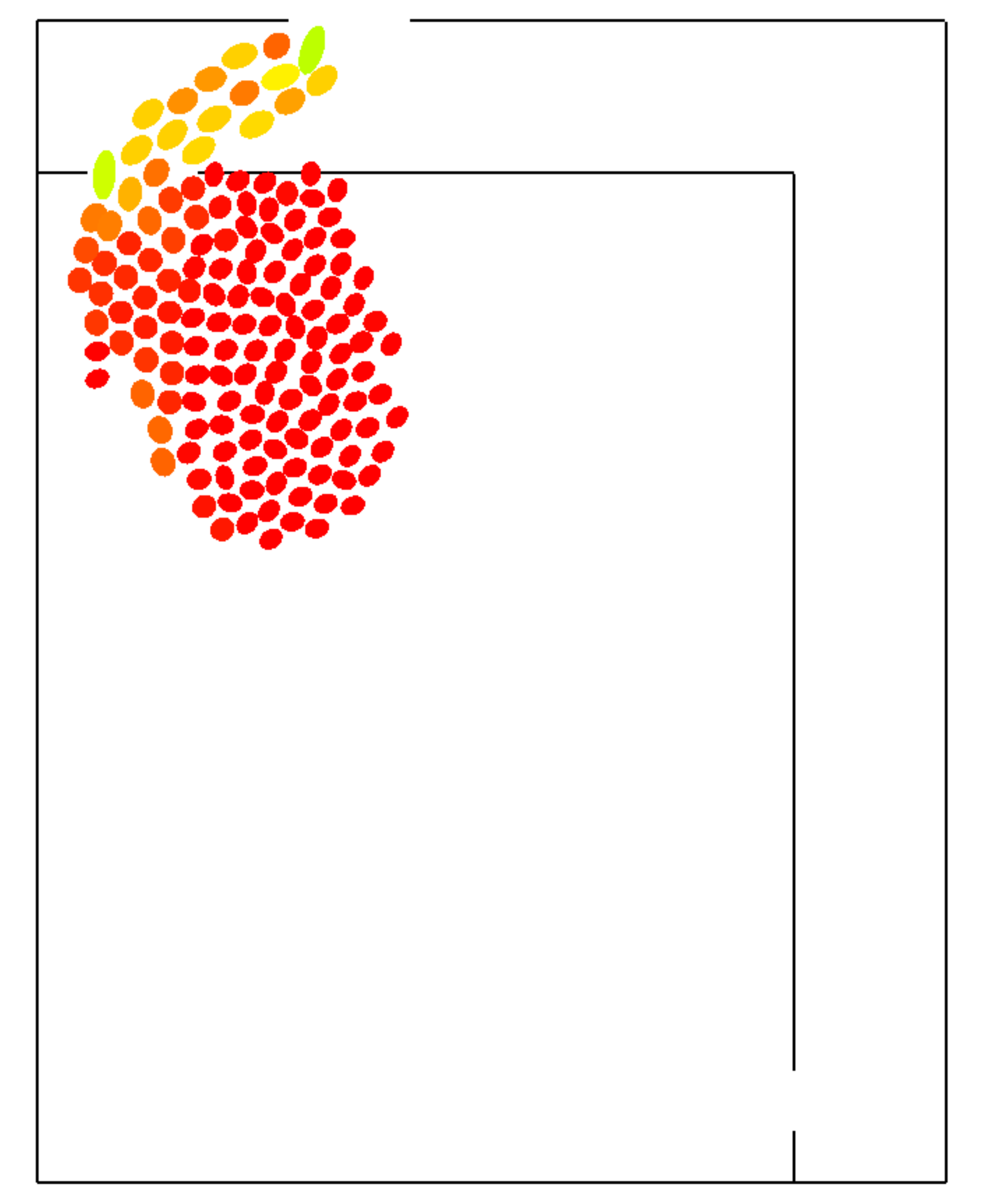}}
\qquad
\subfloat[Global shortest with quickest path]{\label{fig:GSQ_DEMO}\includegraphics[width=35mm]{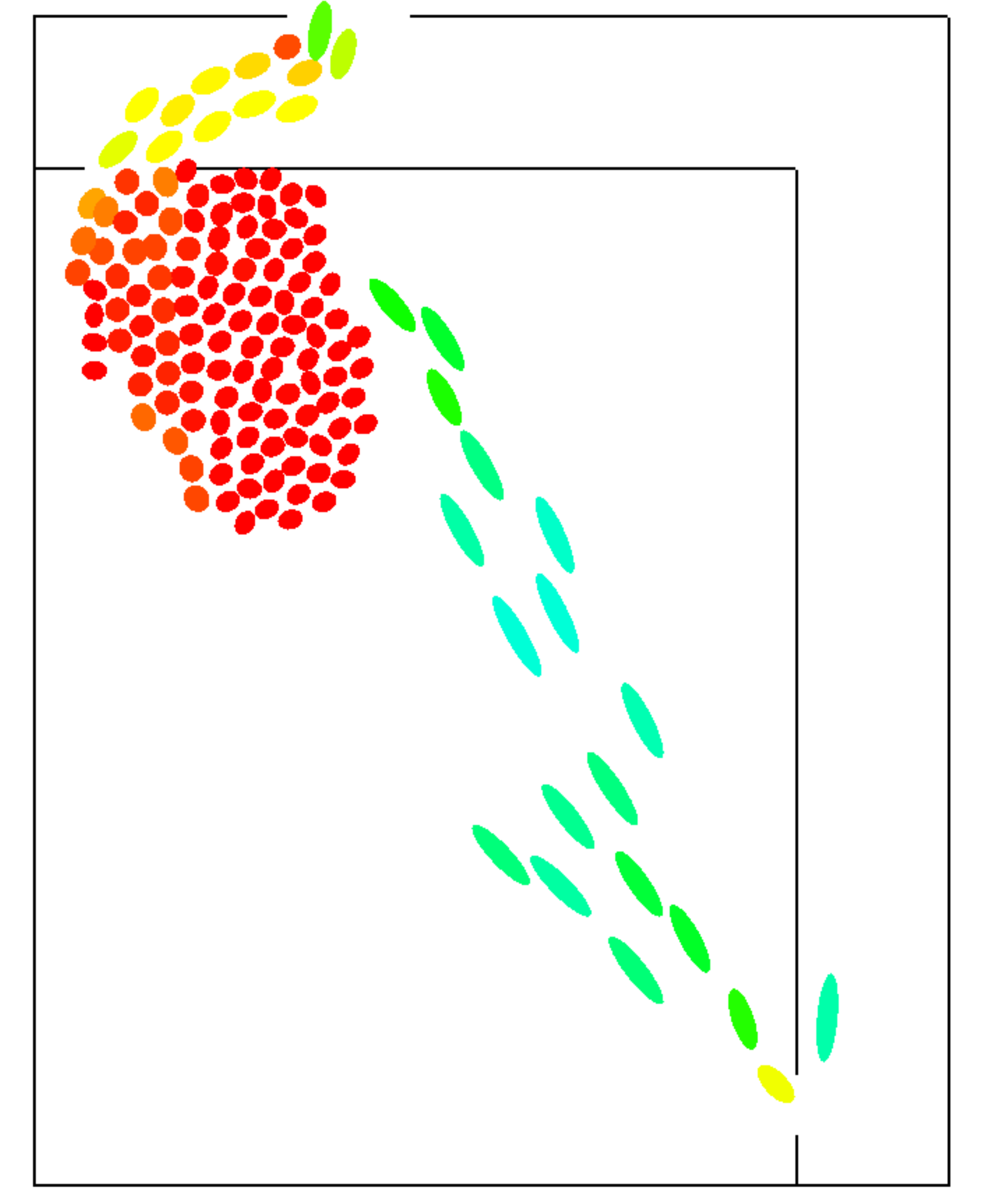}}
\caption[]{Dynamics of the system after 60 seconds. Congestions areas are red. In \subref{fig:LSP_DEMO} pedestrians follow the local nearest exit. In \subref{fig:GSP_DEMO} 
the global nearest exit is chosen. In  \subref{fig:LSQ_DEMO} and  \subref{fig:GSQ_DEMO} pedestrians are allowed to change their current destination.}%
\label{fig:evacuation_time_demo}%
\end{figure}

\subsection{Evacuation time}
The evacuation time is given by the last person leaving the facility. Another definition can be a clearance of the building up to 95\% of the occupants. 
We consider the former definition here. 
A visual assessment of the evacuation dynamics for a simple scenario  after 60 seconds is given by Fig. \ref{fig:evacuation_time_demo}. 
The colour of the pedestrians is correlated to their current velocity. Red means slow  and represents congestions areas. 
Green corresponds to the maximum desired velocity of the pedestrians. The desired velocities are Gaussian distributed 
with mean $1.34\:ms^{-1}$ and standard deviation $0.26\:ms^{-1}$. The evacuation times distribution is presented in 
Fig. \ref{fig:evacuation_time_200_peds}. The smallest variance is achieved by the GSP. This is quite normal as the GSP remains 
the same for all pedestrians independent of their initial positions. The largest variance is given by the LSP. 
The dynamics brought by the quickest path leads not only to a reduction of the overall evacuation time and  a reduction of 
the width but also to realistic shapes in the simulation.

\subsection{Jamming time}
\begin{figure}[htb]
\centering
\subfloat[Evacuation time]{
	\label{fig:evacuation_time_200_peds}
	\includegraphics[height=70mm]{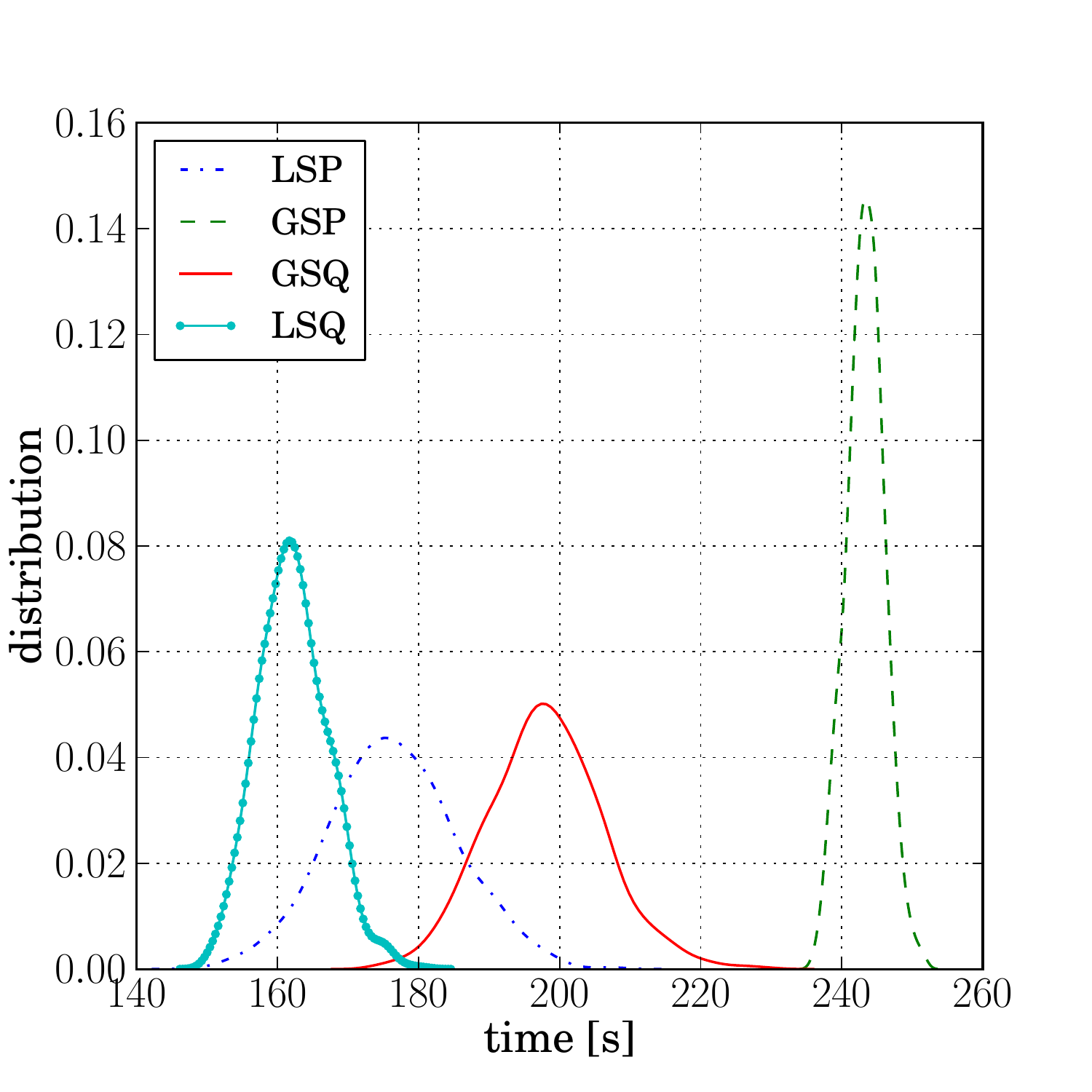}
}
\subfloat[Time in jam]{
	\label{fig:jamming_time_distribution_200_peds}
	\includegraphics[height=70mm]{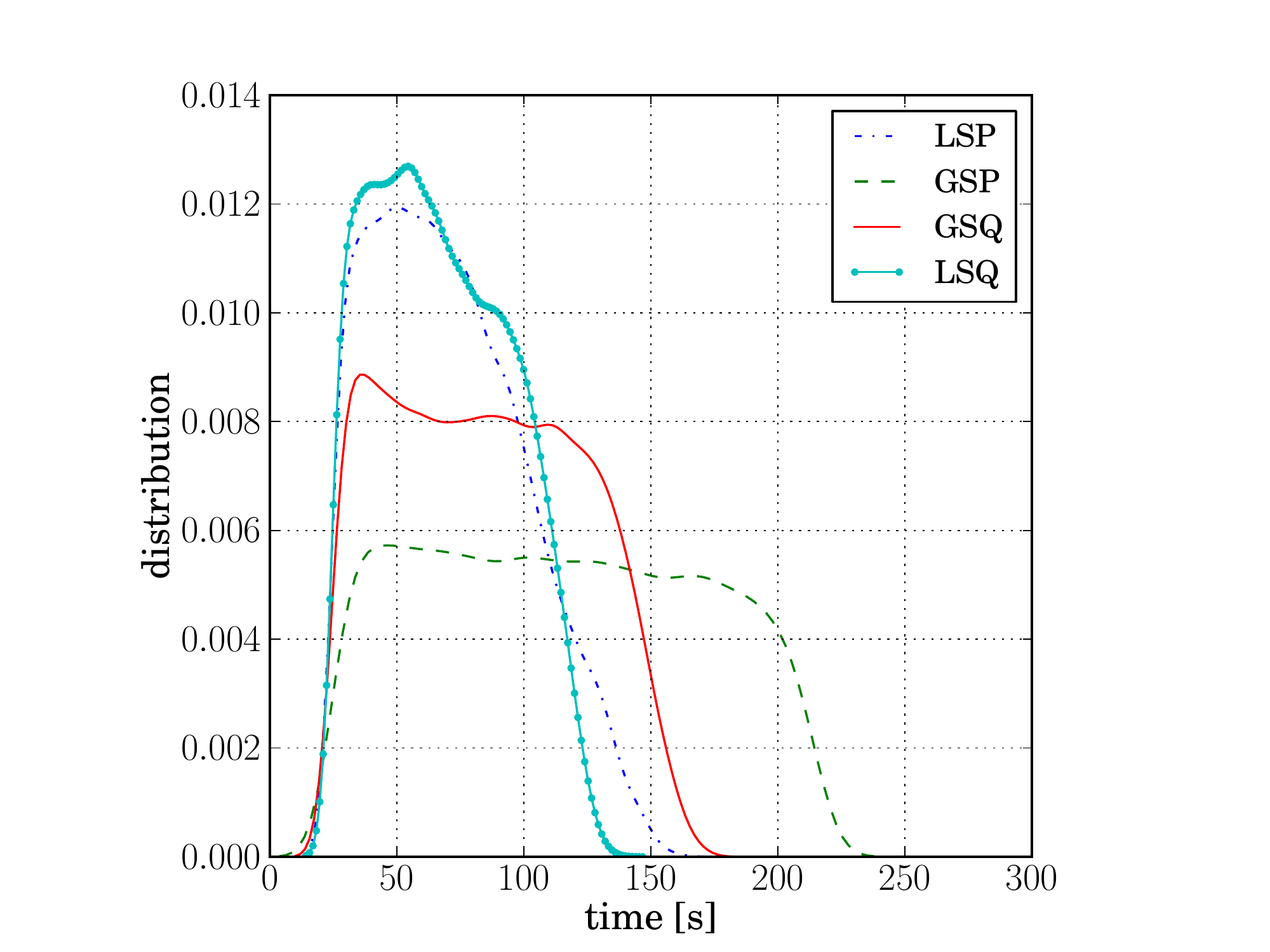}
}
\caption[]{Evacuation time  and time in jam distribution for 200 pedestrians in the scenario in Fig. \ref{fig:evacuation_time_demo}}%
\label{fig:abe}%
\end{figure}

Up to now little importance has been brought to jam analysis itself i.e. how long pedestrian stay in jam.
 This is strongly coupled with the implemented routing strategy. The keyword jam is unfortunately not well defined 
in pedestrian dynamics. A rather crude definition would be to have an absolute zero velocity over a minimal time interval. 
The minimal time interval is needed to avoid very short velocity reduction at sharp turn for instance. 
We consider pedestrians moving at a speed lower than $0.2\:ms^{-1}$ \cite{seyfried2010b} for a period of at least 10 seconds as being in a jam situation. 
The total time in jam is recorded for each pedestrian and a distribution of the recorded times is calculated.  The jamming time distribution for 
 the simulation scenario given by Fig. \ref{fig:evacuation_time_demo} is presented in Fig. \ref{fig:jamming_time_distribution_200_peds}.
As expected the width of the distribution is smaller for the quickest path.

\subsection{Jam size}
\begin{figure}[htb]
\centering
\subfloat[Initial homogeneous distribution]{
	\label{fig:jam_evolution_initial_distribution}
	\includegraphics[height=60mm]{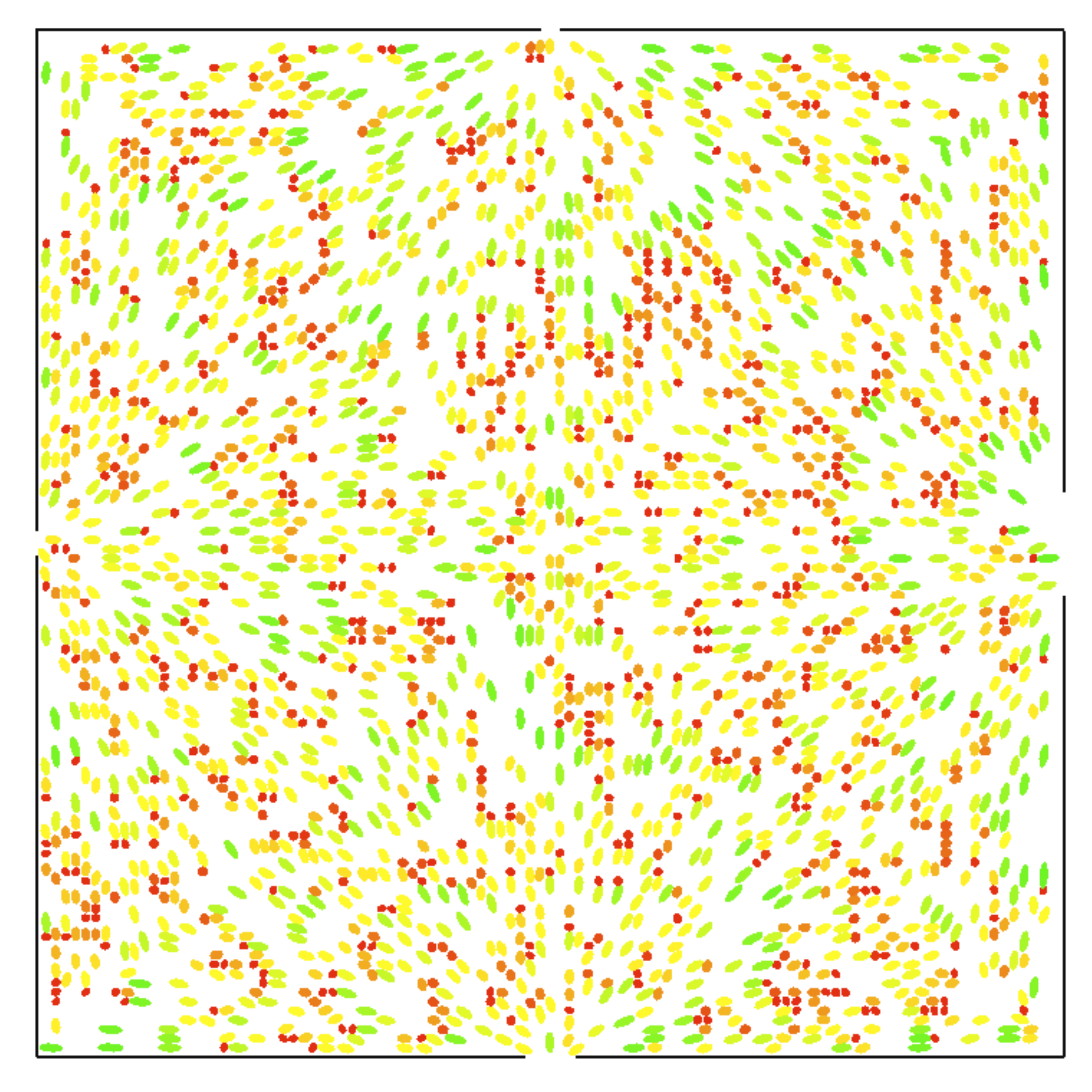}
}
\subfloat[Jam size evolution at the four exits for the local shortest path combined with the quickest path.]{
	\label{fig:jam_size_evolution_initial_distribution}
	\includegraphics[height=65mm]{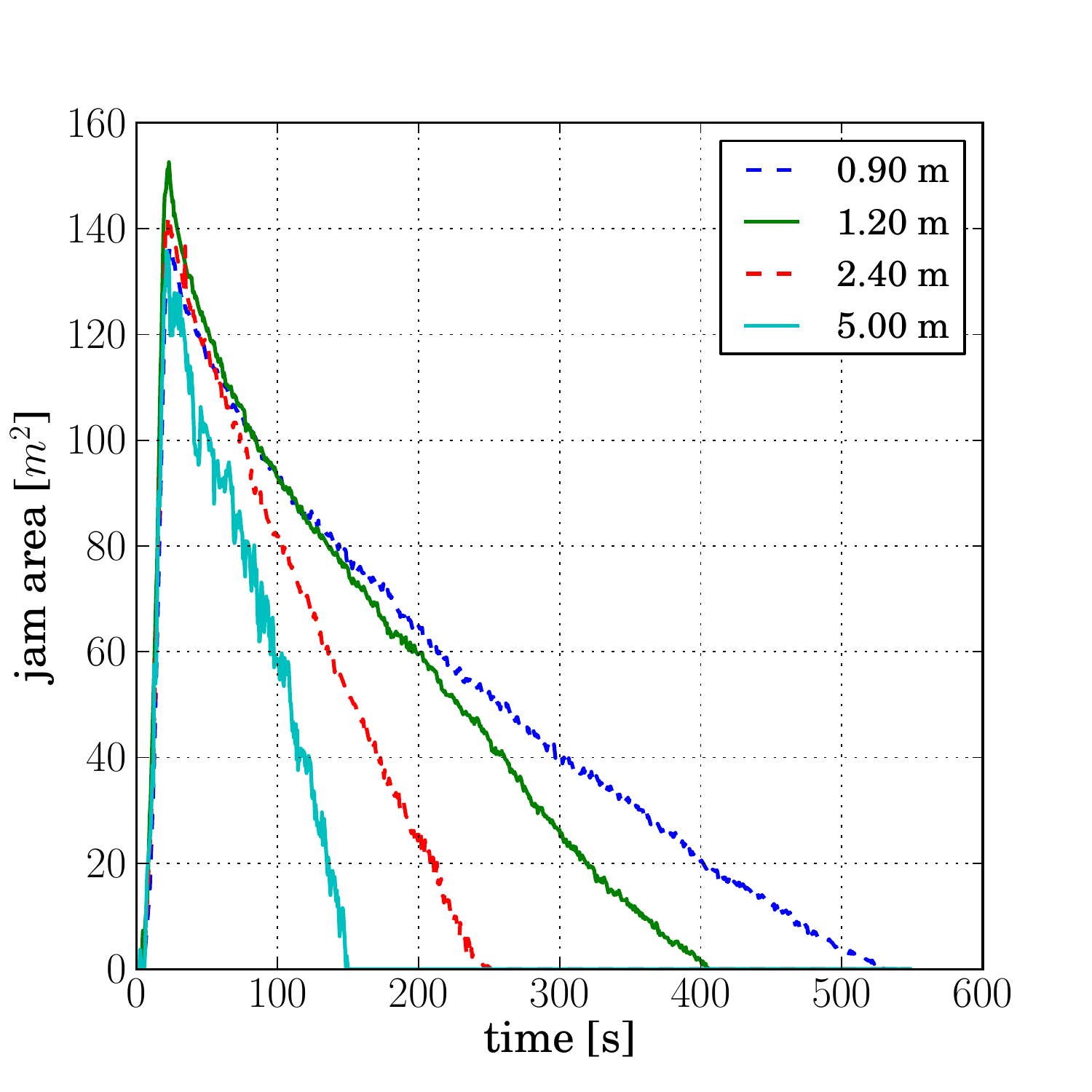}
}

\caption[]{Initial distribution and jam size evolution of 2500 pedestrians in a 50 $m$ x 50 $m$ 
			room with four exits on each wall side. The exits width are 90 $cm$, 120 $cm$, 240 $cm$ and 500 $cm$.}%
\label{fig:jam_size_initial_distribution}%
\end{figure}

The jam size and its evolution cannot directly be derived from the evacuation time and/or the individual times spent in jam. 
It has to be analysed separately and for each exit individually. Short-lived jams are rather uncritical, the same holds for  short-sized jams.
 We are particularly interested in big jams with a long lifetime. They can be  dangerous and should be 
checked against the characteristics of the evacuees for instance their ages, but also against the environmental conditions, 
the temperature for instance.
The jam size is calculated by summing up the effective areas occupied by the pedestrians in jam at each time step. 
This is easily done by summing up the ellipses areas representing the pedestrians in the GCFM. Another method could be to build the envelope of the pedestrians
and calculate the area of the resulting polygon. Fig. \ref{fig:jam_evolution_initial_distribution} shows a 
straight forward example of a jam size analysis scenario after the first time step (0.01 second). The colour corresponds to the states. 
Red pedestrians are waiting for a possibility to move. Other pedestrians are already moving. The initial density is on average 1 $pedestrian/m^2$. 
The corresponding average jam areas are presented in Table \ref{tab:areas_demo}. The values for the global and local shortest path are the same. 
There as some  difference in their corresponding combination with the quickest path. 
At a first look the values for the different routing strategies do not differ  much from each other, but one has to consider 
the difference in the evacuation time, which is more than one minute in this case. There is a distribution of 
the overall load from the congested exits to the non congested ones.\\
The  jam size evolution at the different exits using the quickest path is shown in Fig. \ref{fig:jam_size_evolution_initial_distribution}. 
There is a sharp increase at the beginning of the simulation. Peaks of approximately 150 $m^2$ are reached at all exits. 
The values then decrease with a slope correlated with the routing strategy used. The slope is constant with the static strategies. The quickest path shows a less constant slope due to the fact that some pedestrians are changing their destinations.

\begin{table}[htb] 
\caption{Average jam size at different exits  of Fig. \ref{fig:jam_size_initial_distribution} for the four strategies.} 
\centering  
\begin{tabular}{c|cccc}  
\hline\hline
\multirow{2}{*}{Door width $[m]$}   
& \multicolumn{4}{c}{Jam area $[m^2]$}\\ \cline{2-5}
& LSP & GSP &GSQ  & LSQ\\ 
\hline  
0.90 & 53.7  & 53.7 & 48.8 & 49.3\\
1.20 & 52.8 & 52.8 & 48.2 & 42.5\\
2.40 & 27.4 & 27.4 & 31.3 & 28.1\\
5.00 & 13.2 & 13.2  & 20.3 & 18.1\\
\hline \hline
\end{tabular} 
\label{tab:areas_demo} 
\end{table}

\section{Simulation and Analysis}

\begin{figure}[htb]
\centering
\subfloat[250 pedestrians distributed in a single block.]{\label{fig:arena_single}\includegraphics[width=70mm]{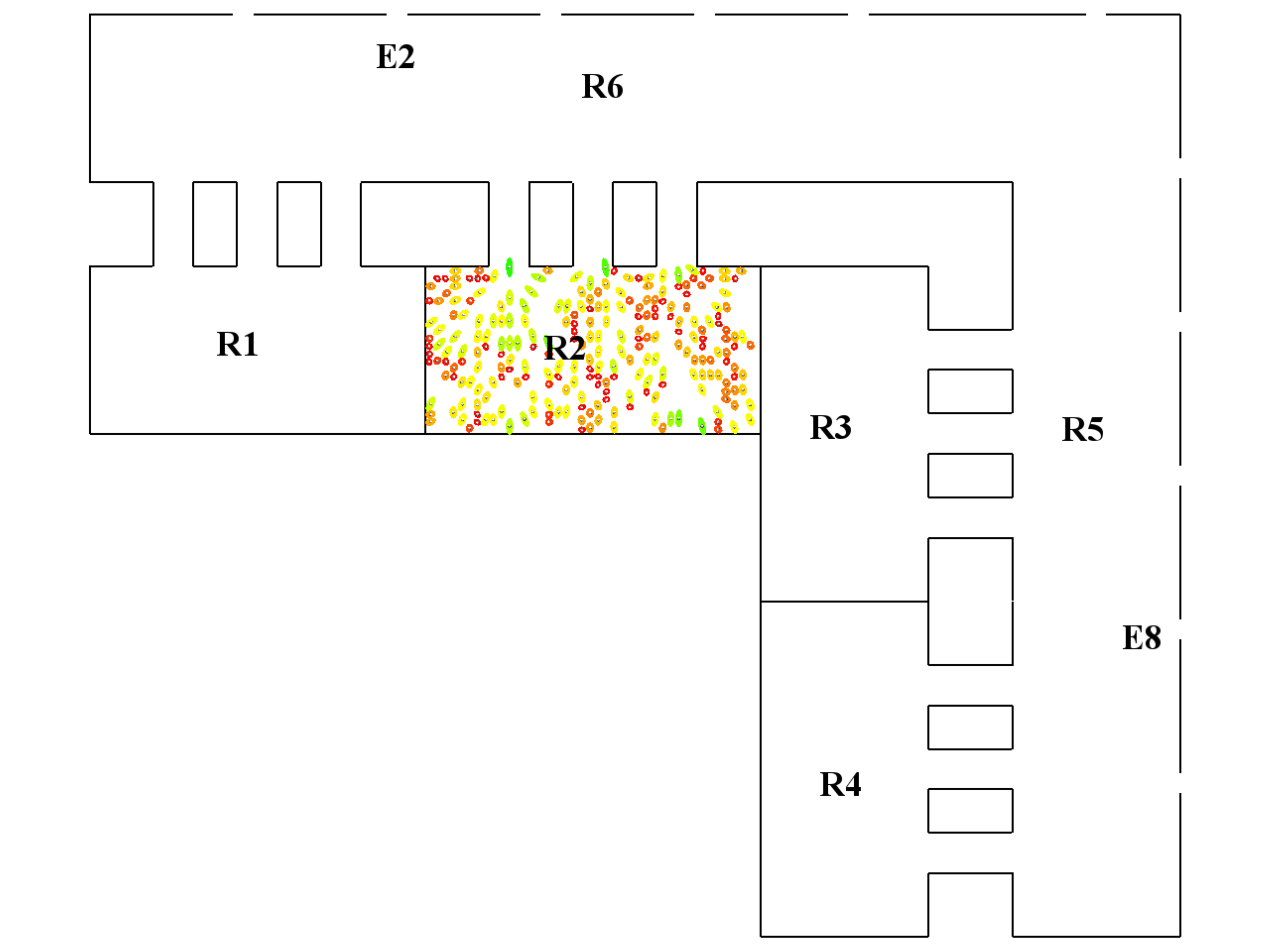}}
\subfloat[1000 pedestrians distributed in four blocks.]{\label{fig:arena_complete}\includegraphics[width=70mm]{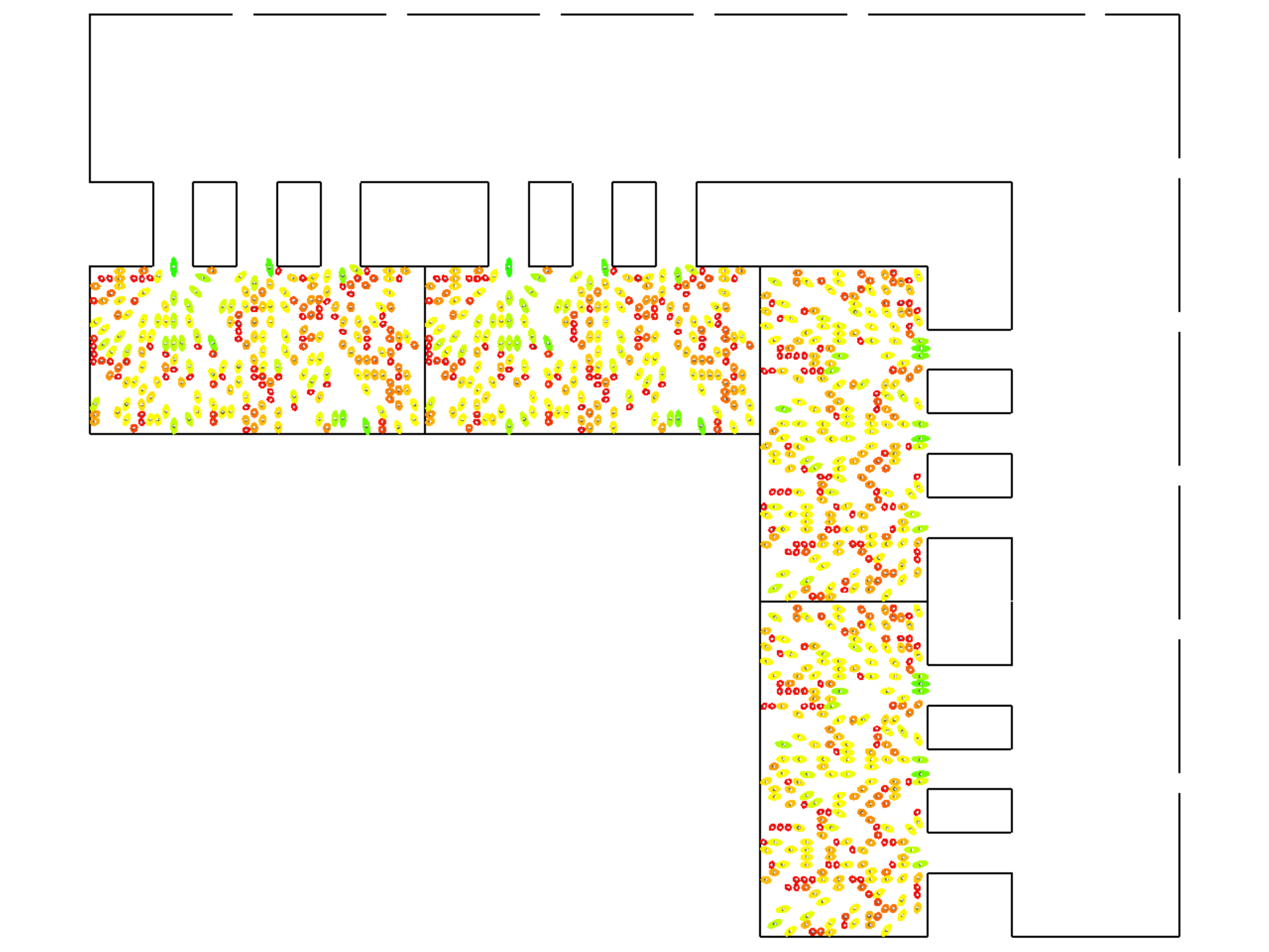}}
\caption{Investigated facility with initial homogeneous distributions of pedestrians. The total simulation area is 2144 $m^2$.}%
\label{fig:initial_distribution}%
\end{figure}

The strategies discussed in the previous section are tested on the structure described in Fig. \ref{fig:initial_distribution} 
with different initial distributions. The simulation area is a simplified model of a section of the ESPRIT arena in D\"usseldorf, 
Germany which holds up to 50.000 spectators. The room R1 to R4 are the grandstand with dimensions 10  x 20 $m^2$. 
The tunnels are 2.40 $m$ wide and 5 $m$ long. The exits are 1.20 $m$ wide. The rooms R4 and R6 are 10 $m$ wide. 
The criteria used to evaluate the simulation results are presented in the previous section. 

\begin{figure}[htb]
\centering
\subfloat[Local shortest path]{\label{fig:LSP_single}\includegraphics[width=50mm]{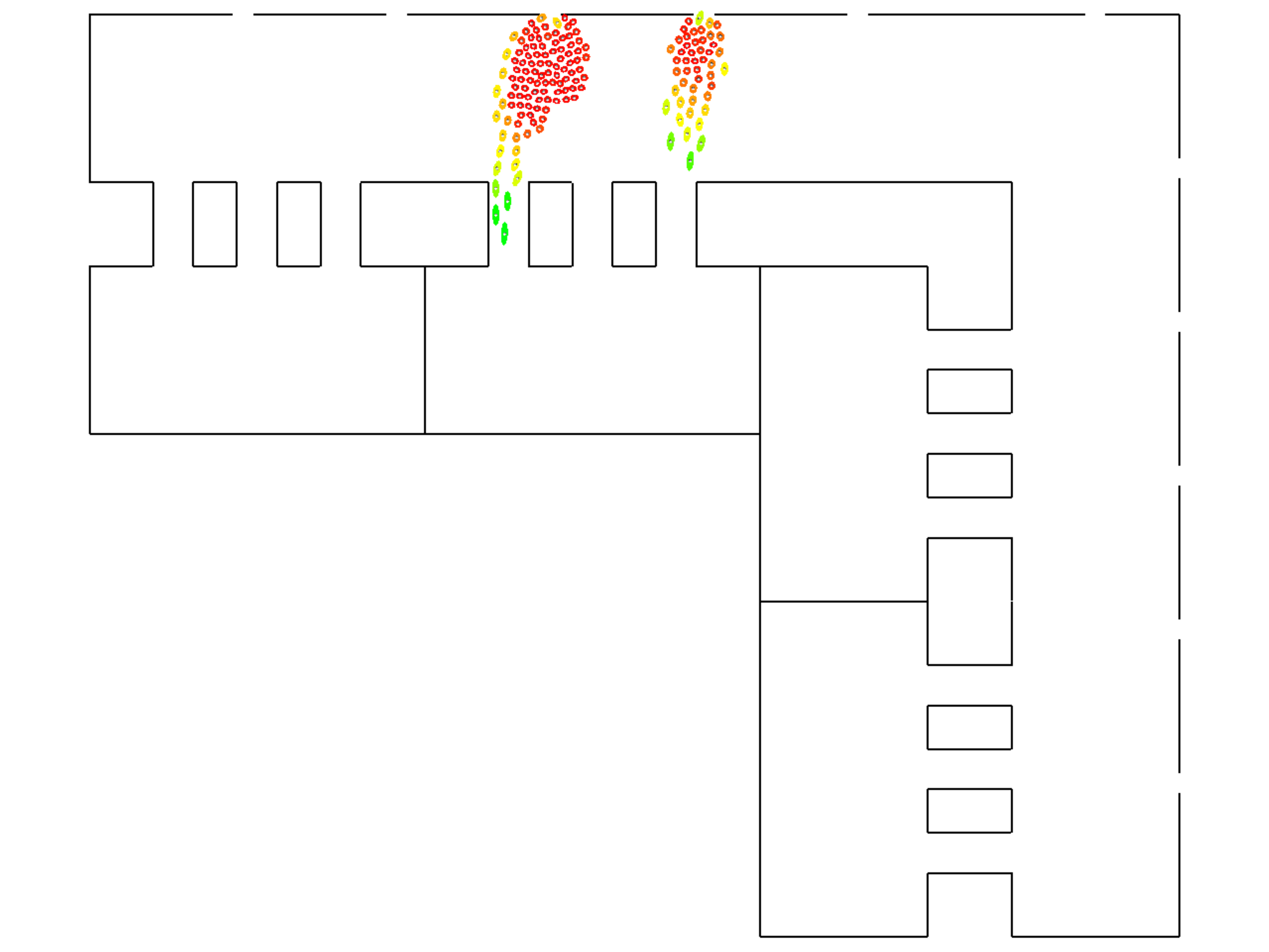}}
\subfloat[Local shortest with quickest path]{\label{fig:LSQ_single}\includegraphics[width=50mm]{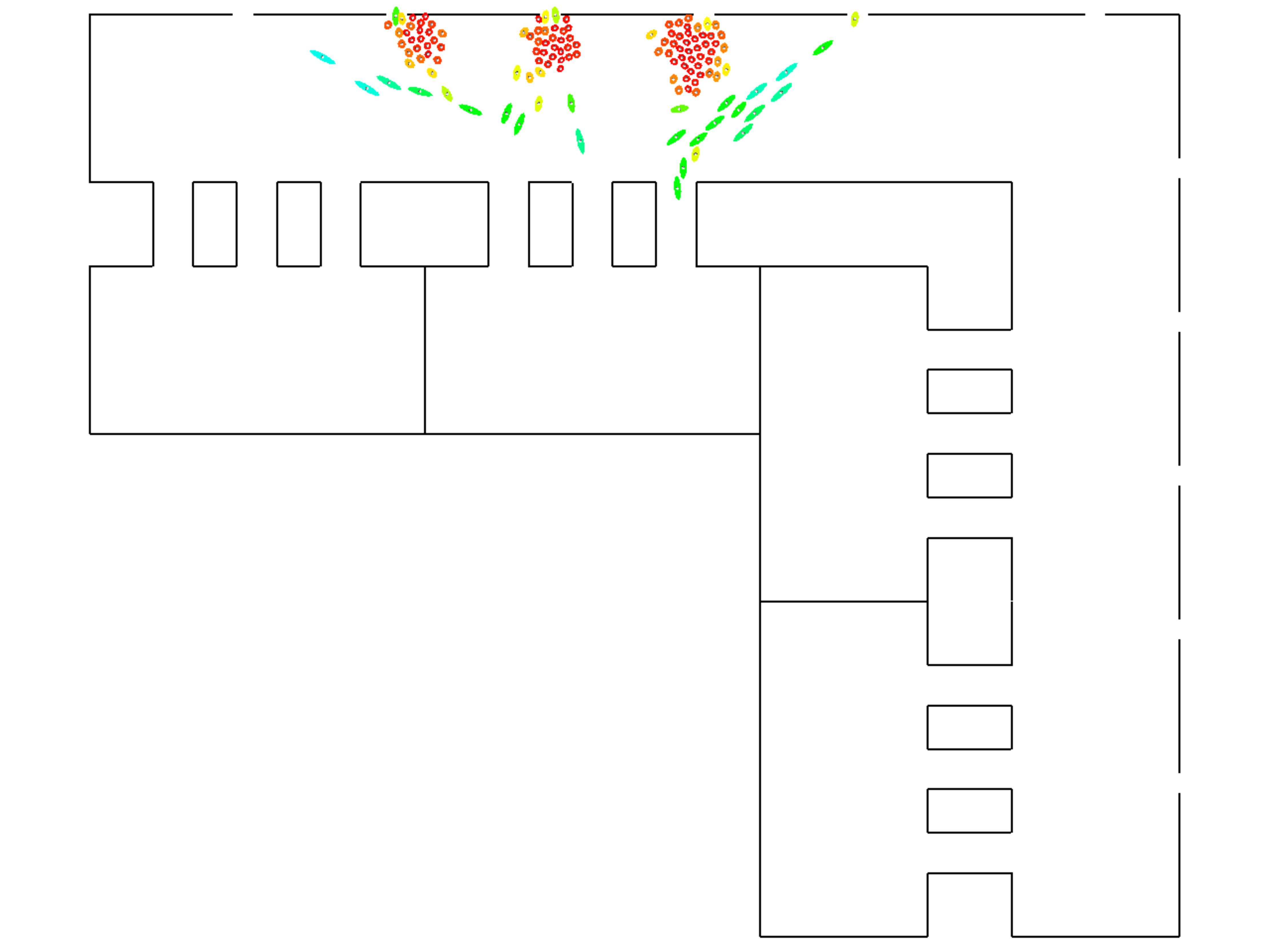}}\\
\subfloat[Global shortest path]{\label{fig:GSP_single}\includegraphics[width=50mm]{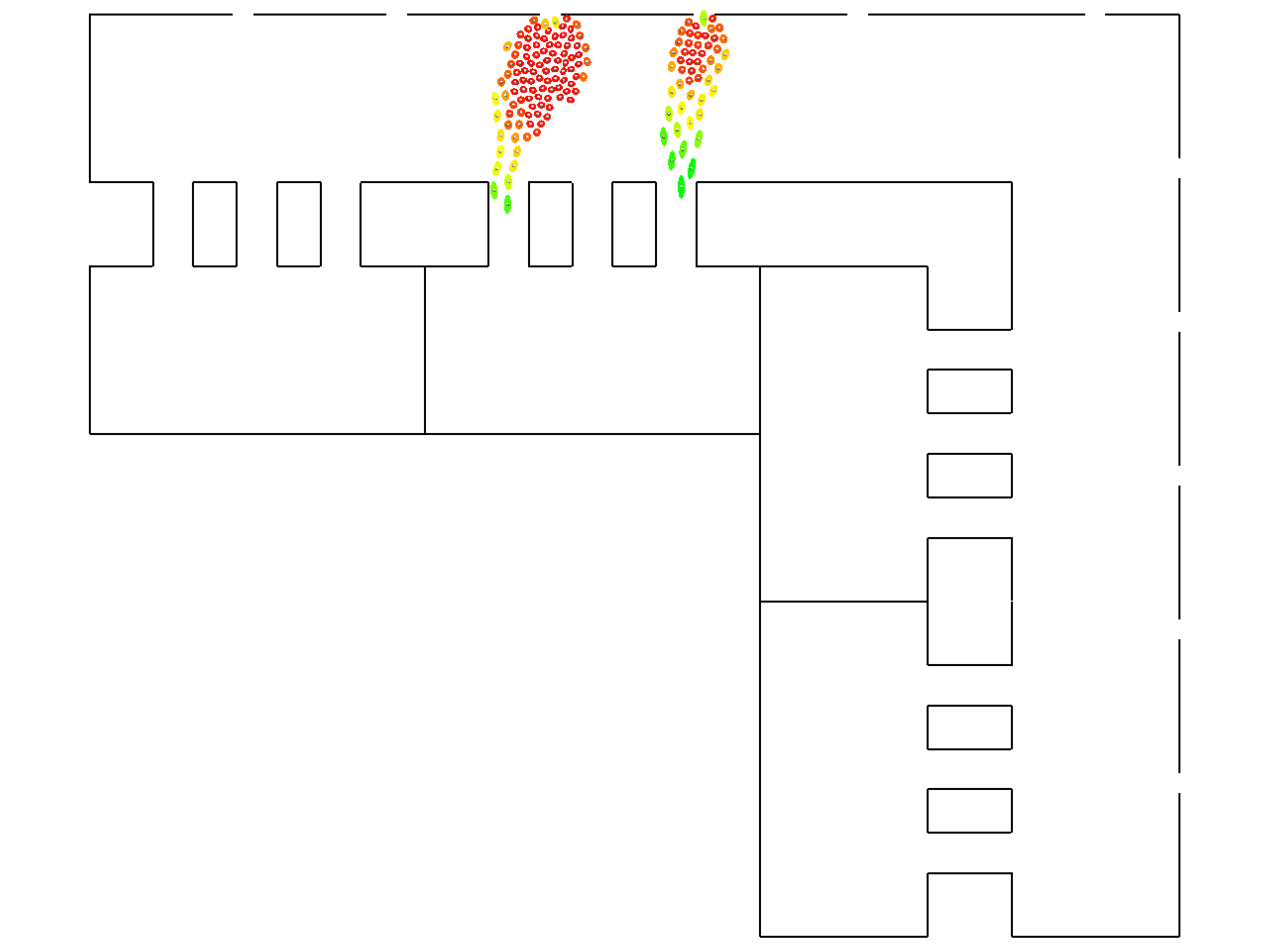}}
\subfloat[Global shortest with quickest path]{\label{fig:GSQ_single}\includegraphics[width=50mm]{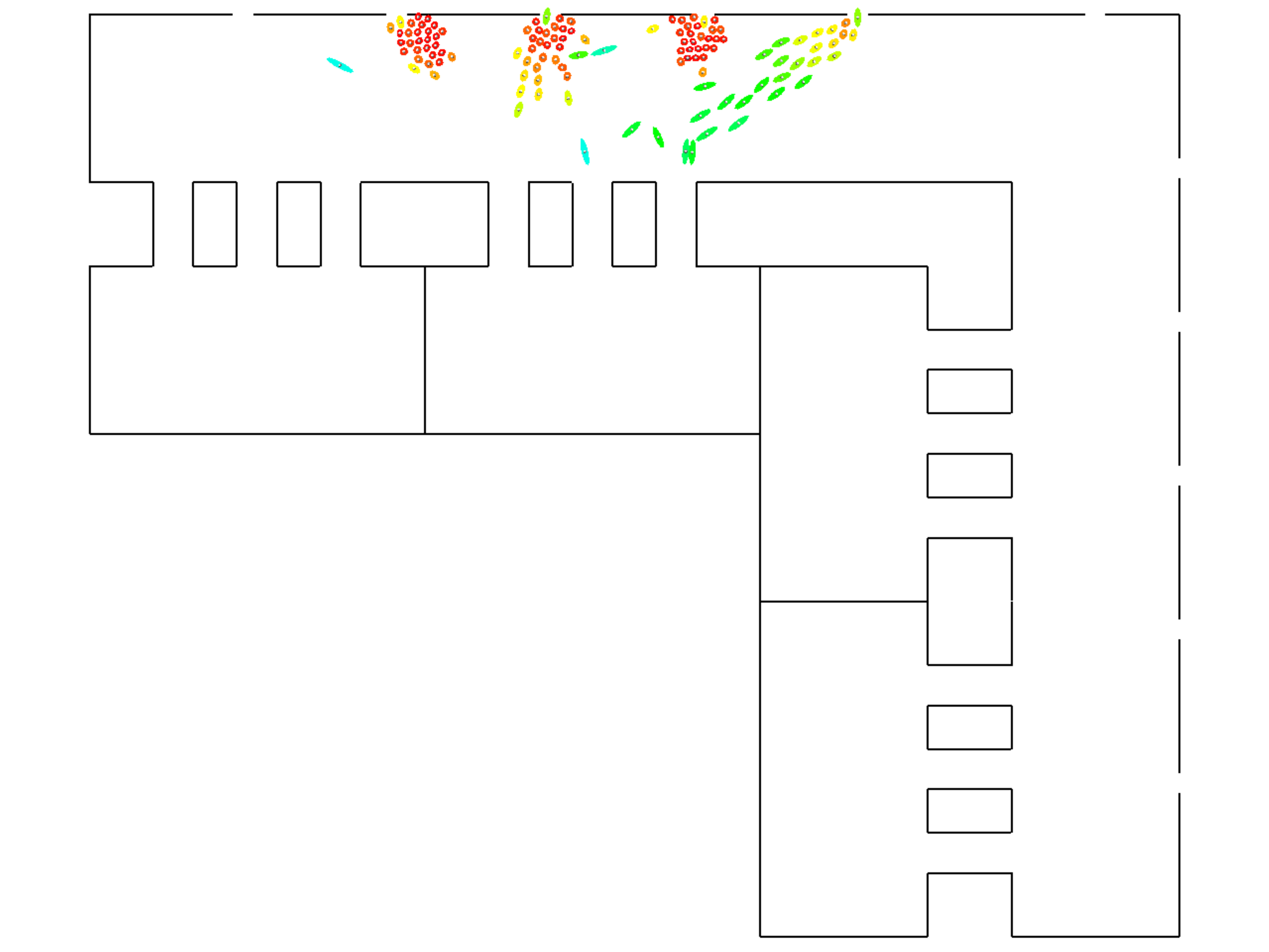}}
\caption[]{Dynamics of the system after 60 seconds for the initial distribution in Fig. \ref{fig:arena_single}. Congestions areas are red.}%
\label{fig:dynamic_single}%
\end{figure}

In the first simulation case we are interested in the evacuation of a single block of the stadium. Fig. \ref{fig:arena_single} 
shows the initial homogeneous distribution of this test with 250 pedestrians. The dynamic of the evacuation after 60 seconds for 
the four strategies is presented in Fig. \ref{fig:dynamic_single}.  With the shortest path only two exits are effectively used whereas 
the quickest path approach takes advantage of fours exits leading to a faster evacuation.

\begin{figure}[htb]
\centering
\subfloat[Evacuation time distribution]{\label{fig:evac_time_single}\includegraphics[width=50mm]{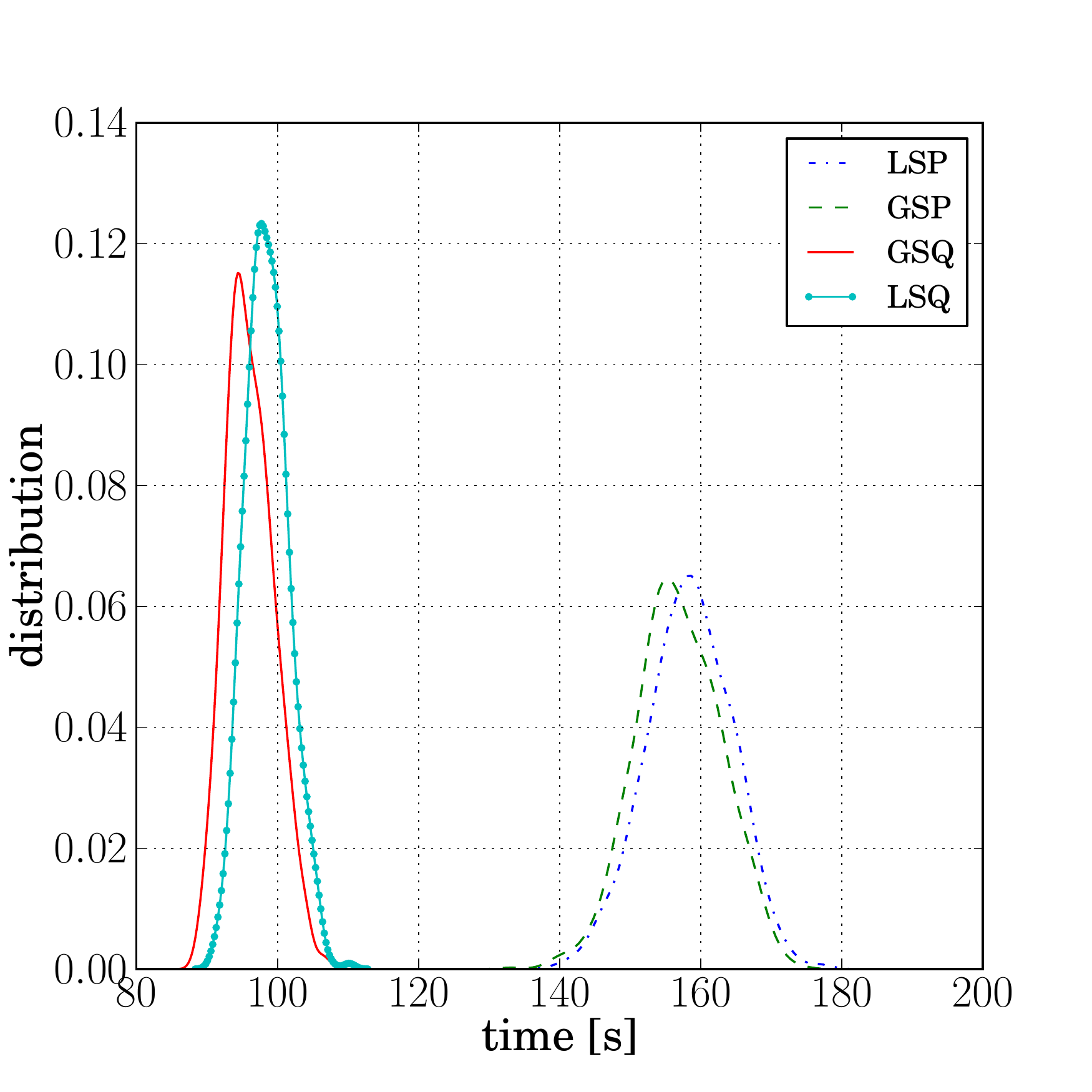}}
\subfloat[Jam time distribution]{\label{fig:jam_time_single}\includegraphics[width=50mm]{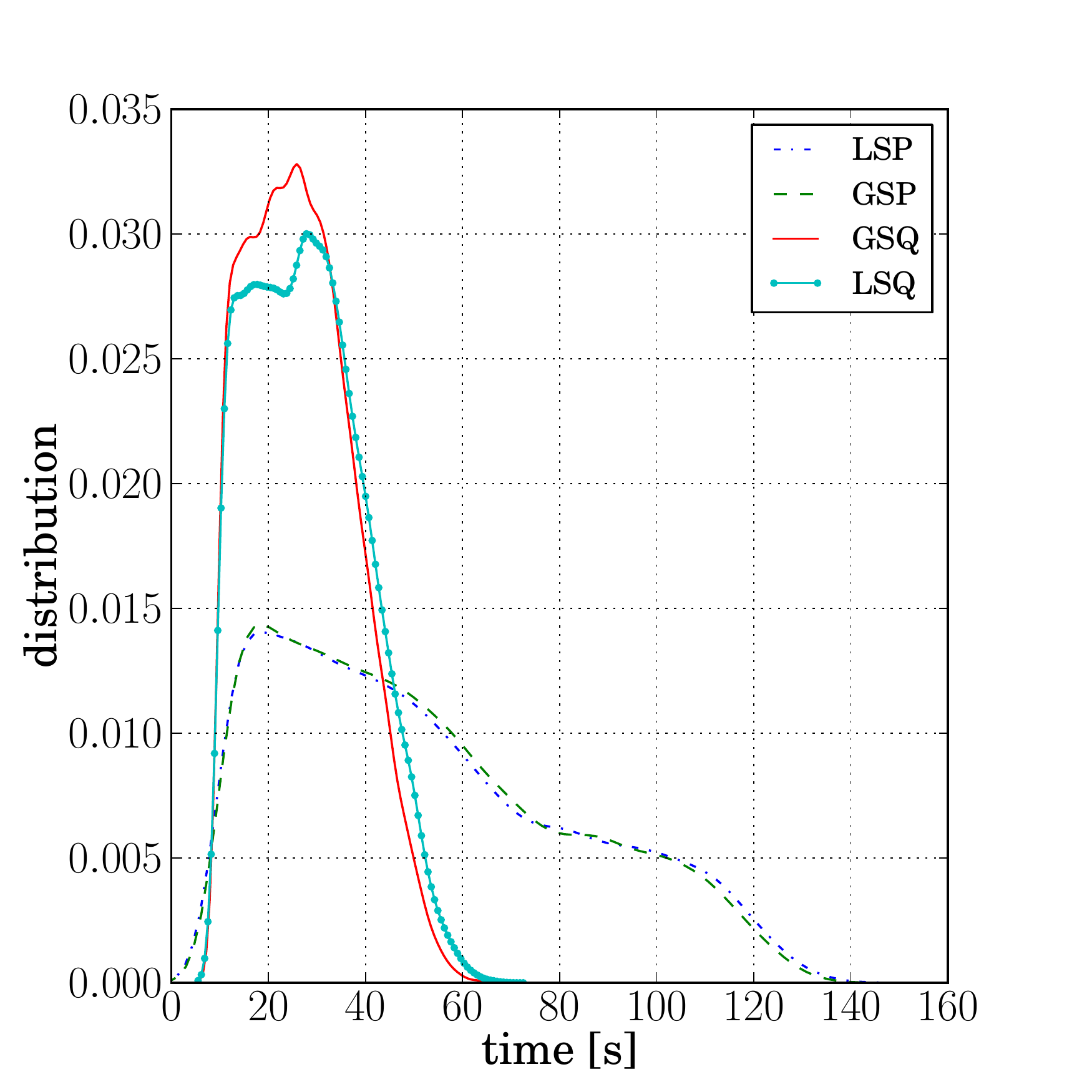}}
\subfloat[Jam size distribution]{\label{fig:250_single_jam_distribution}\includegraphics[width=50mm]{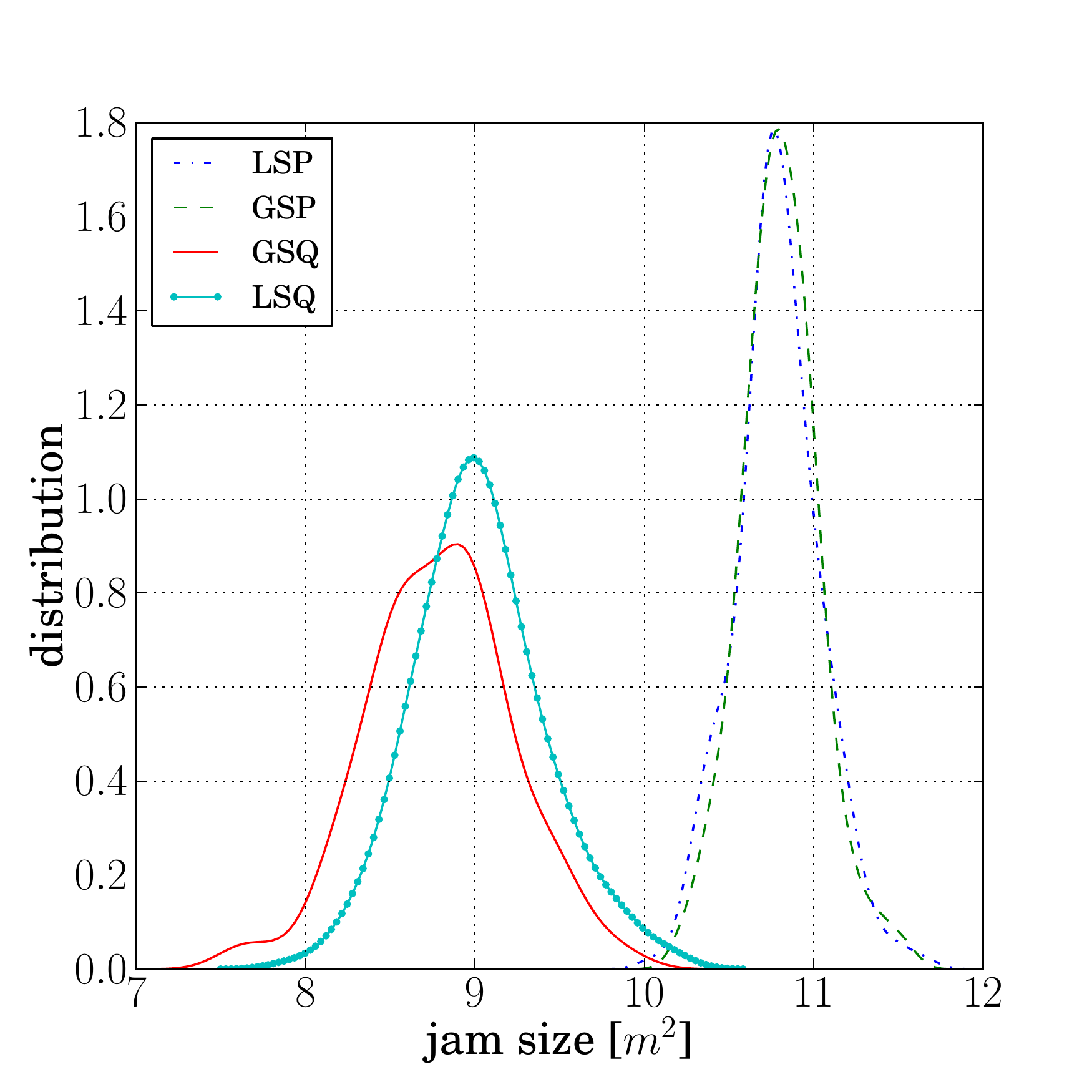}}
\caption{Evacuation time and time in jam distribution for 250 pedestrians. The initial positions are presented in Fig. \ref{fig:arena_single}}%
\label{fig:results_single}%
\end{figure}

\begin{figure}[htb]
\centering
\subfloat[Local shortest path]{\label{fig:LSP_Single_250_jam_evolution}\includegraphics[width=60mm]{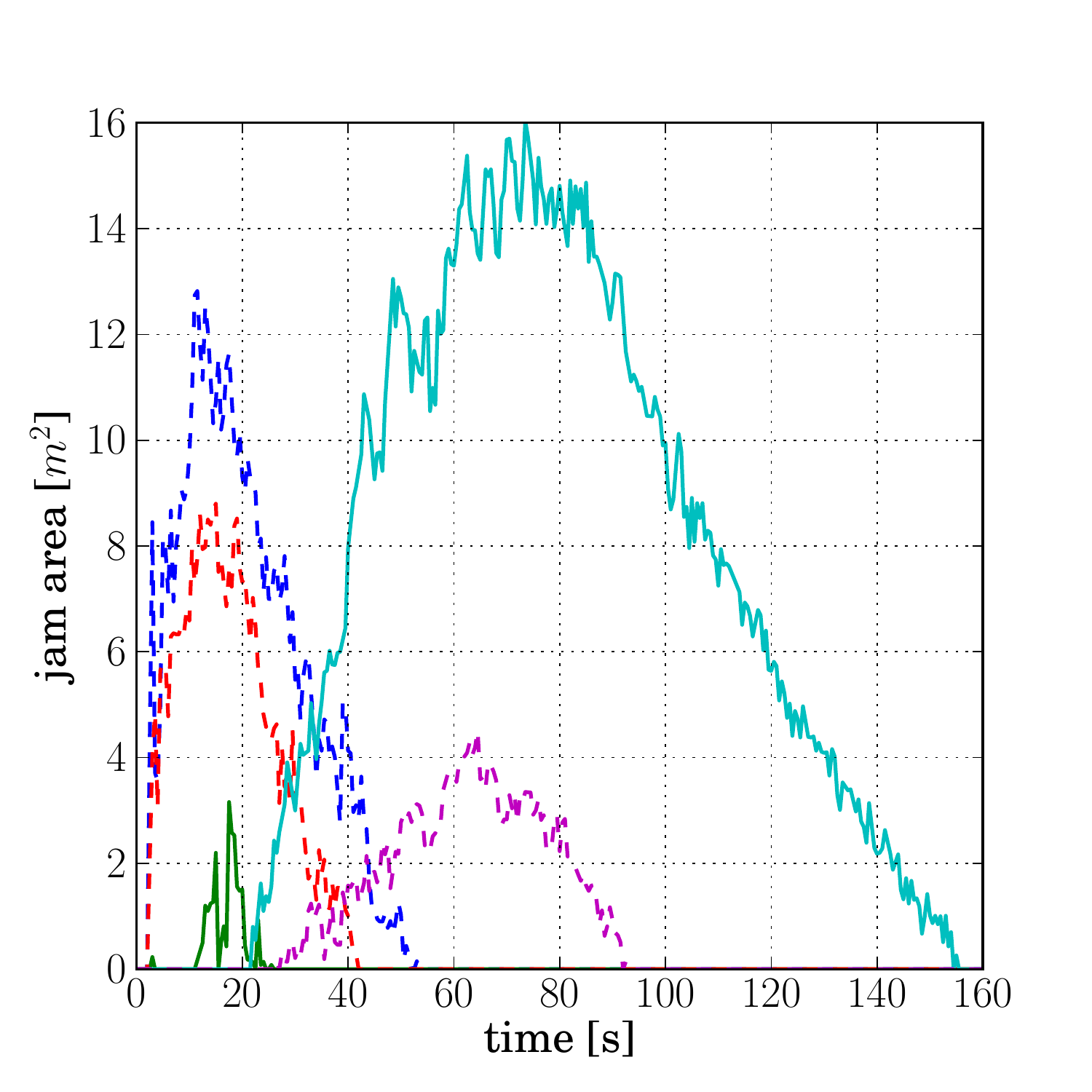}}
\subfloat[Local shortest with quickest path]{\label{fig:LSQ_single_250_jam_evolution}\includegraphics[width=60mm]{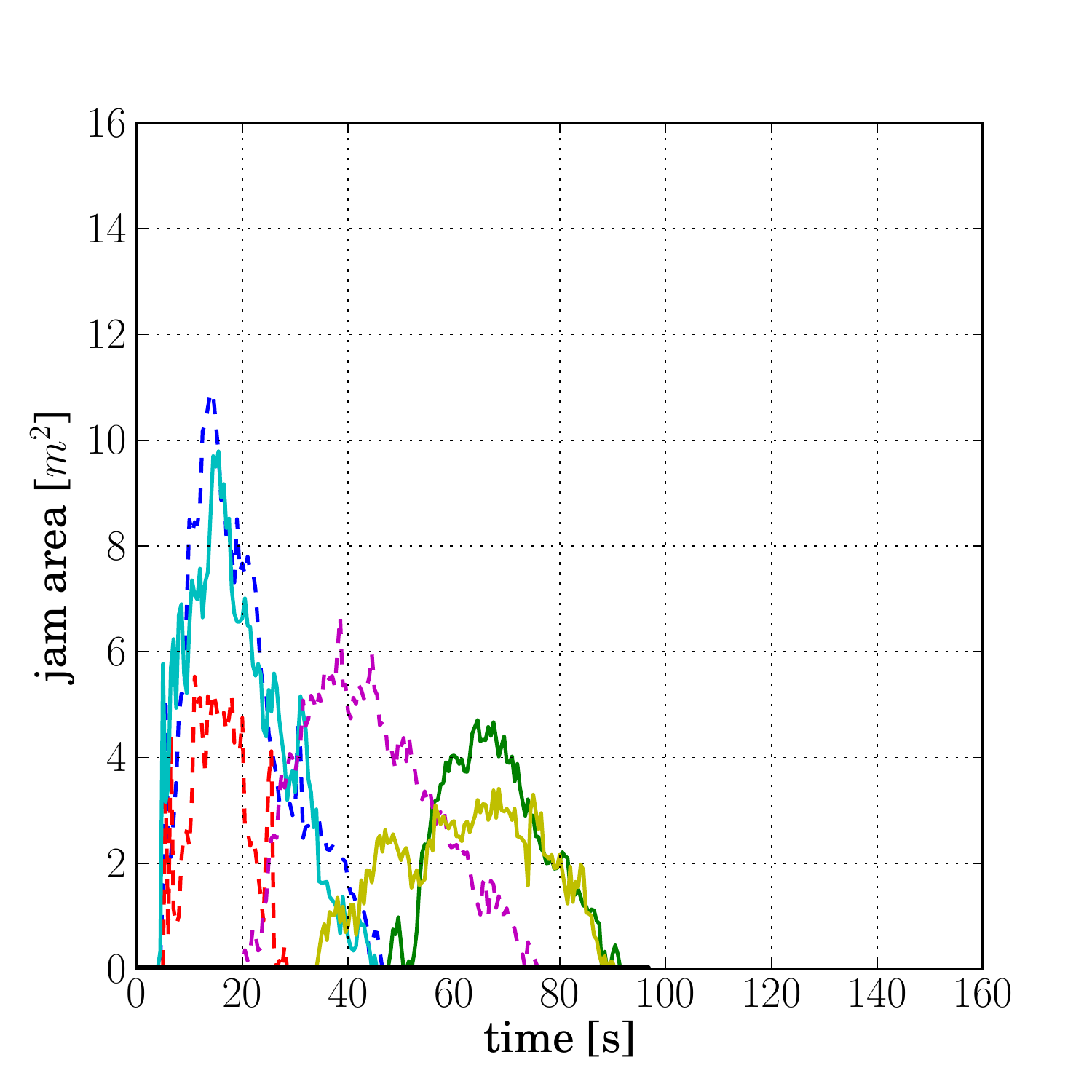}}\\
\subfloat[Global shortest path]{\label{fig:GSP_single_250_jam_evolution}\includegraphics[width=60mm]{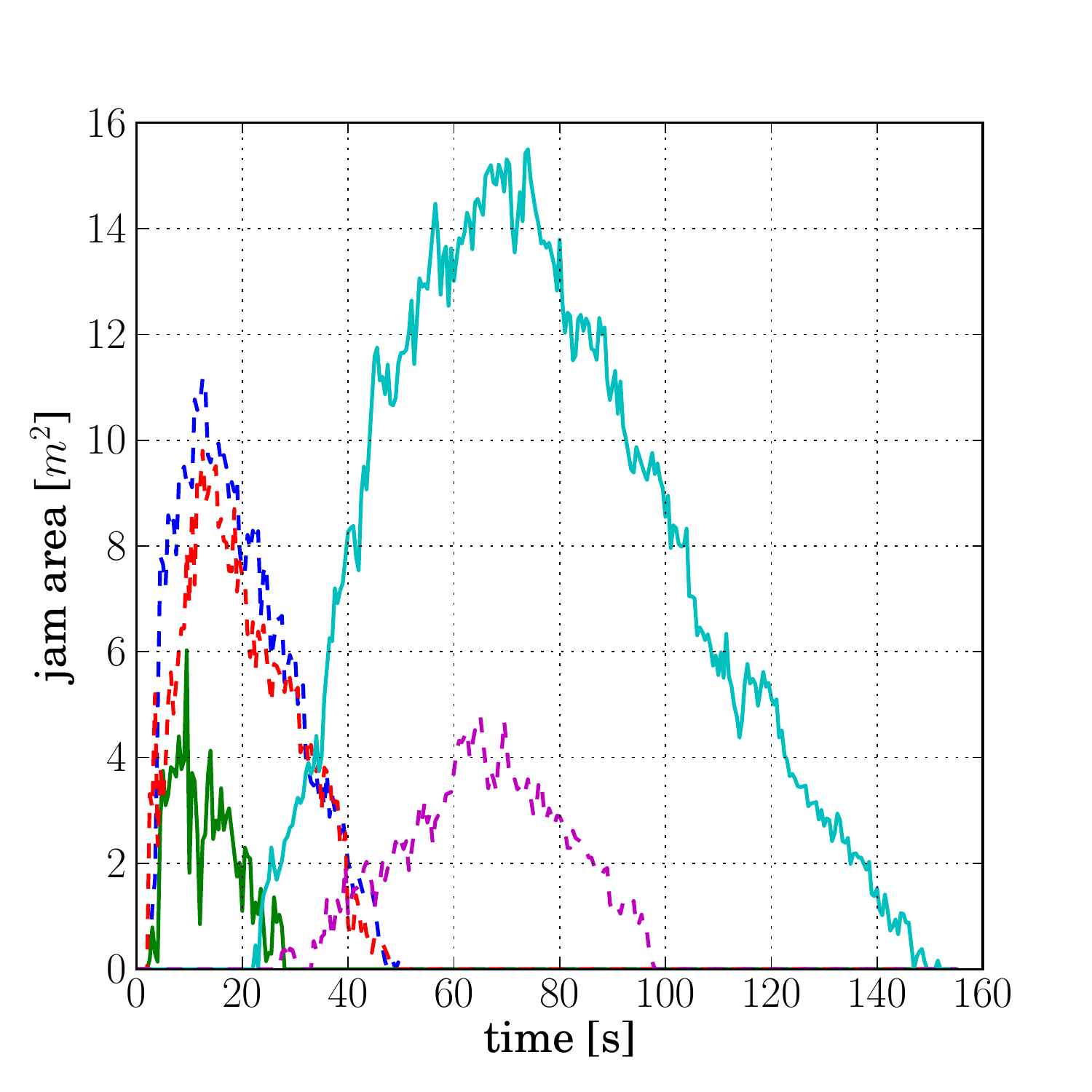}}
\subfloat[Global shortest with quickest path]{\label{fig:GSQ_single_250_jam_evolution}\includegraphics[width=60mm]{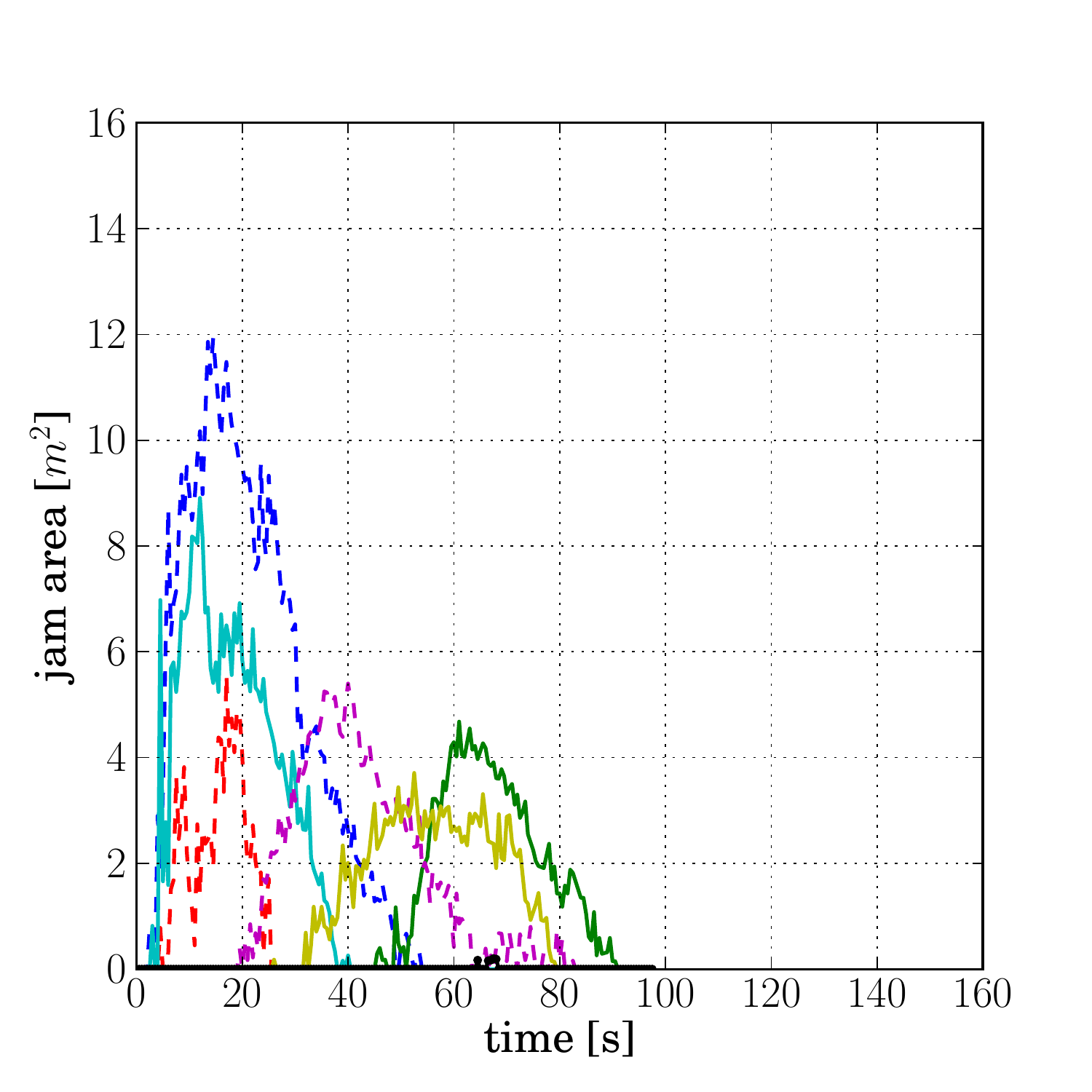}}
\caption[]{Jam size evolution for 250 pedestrians at different exits for the initial distribution in Fig. \ref{fig:arena_single}. The  colours 
correspond to the different exits.
Exits without congestions have been left out of the plots. In \subref{fig:LSQ_single_250_jam_evolution} and \subref{fig:GSQ_single_250_jam_evolution}, there
are  more but short-lived jams than in \subref{fig:LSP_Single_250_jam_evolution} and \subref{fig:GSP_single_250_jam_evolution}.}%
\label{fig:jam_evolution_single}%
\end{figure}

The results of the evacuation time analysis are summarized in Fig. \ref{fig:evac_time_single}.  1000 runs are performed for the evaluation. 
The dynamic brought by the quickest path leads to a faster evacuation, on average 1 minute faster. As depicted in Fig. \ref{fig:dynamic_single}
 other (less congested or free) exits are used, as will be expected in a real evacuation scenario. There is not much difference 
between the results given by the global and the local shortest path, this is due to the shape of the investigated facility. 
The second value analysed is the individual distributions of time 
spent in jam. The results are presented in Fig. \ref{fig:jam_time_single}. As expected the results are correlated to 
the evacuation time. With the quickest path, pedestrians spend less time in jam, 30 seconds  on average. This time is almost tripled using the static strategies. Fig. \ref{fig:jam_evolution_single} shows the evolution of the jam size at different exits.
 Exits without congestions have been left out. The static strategies shows five long lasting jams. The quickest path alternatives 
shows more jams but short lived. The later ones are less critical. The total jam size distribution is taken from Fig. \ref{fig:250_single_jam_distribution}. 
The average jam size  for the static strategies is 11 $m^2$ and 9 $m^2$ for the quickest path.

\begin{figure}[htb]
\centering
\subfloat[Local shortest path]{\label{fig:LSP_complete}\includegraphics[width=50mm]{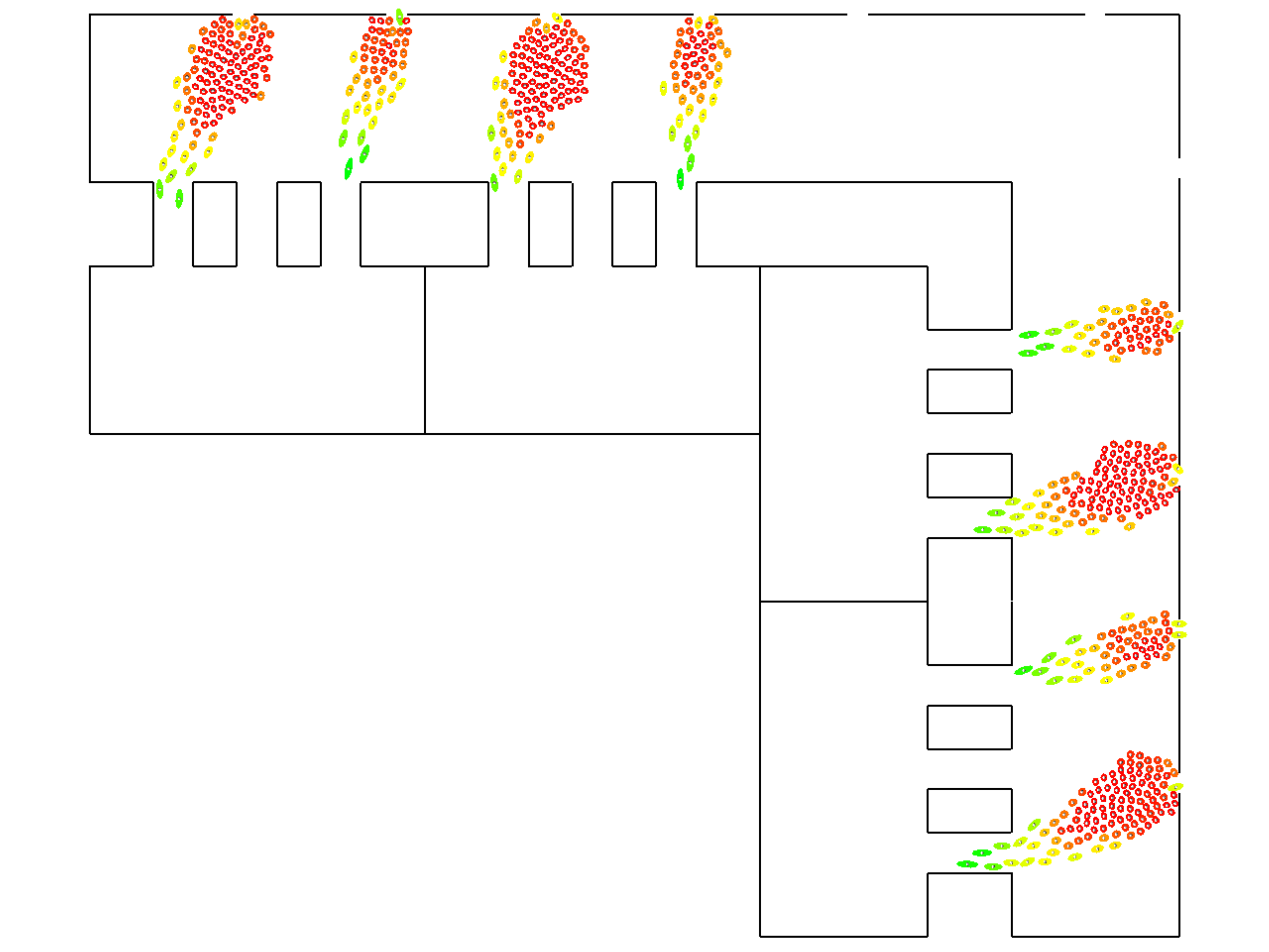}}
\qquad
\subfloat[Local shortest with quickest path]{\label{fig:LSQ_complete}\includegraphics[width=50mm]{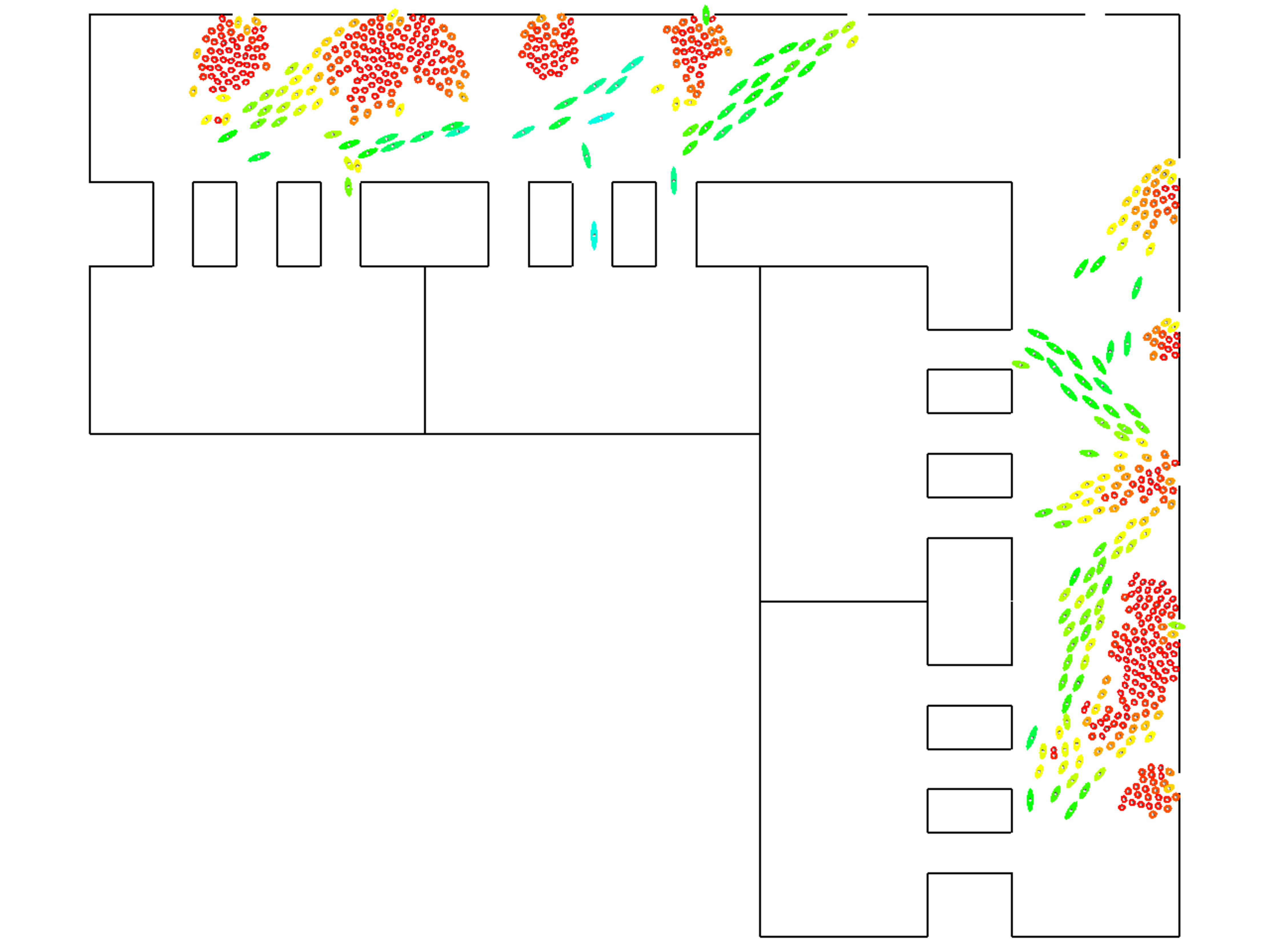}}\\
\subfloat[Global shortest Path]{\label{fig:GSP_complete}\includegraphics[width=50mm]{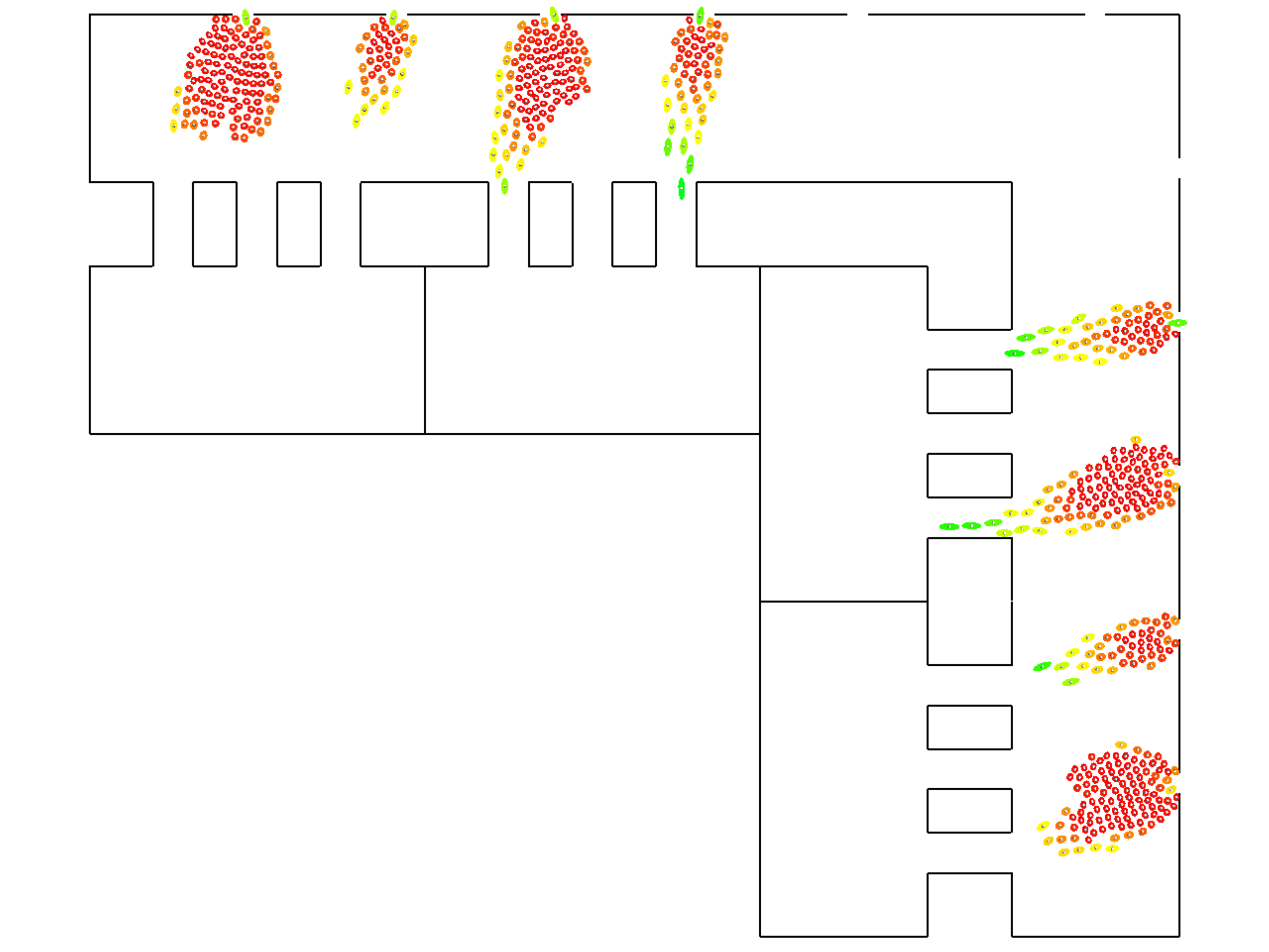}}
\qquad
\subfloat[Global shortest with quickest path]{\label{fig:GSQ_complete}\includegraphics[width=50mm]{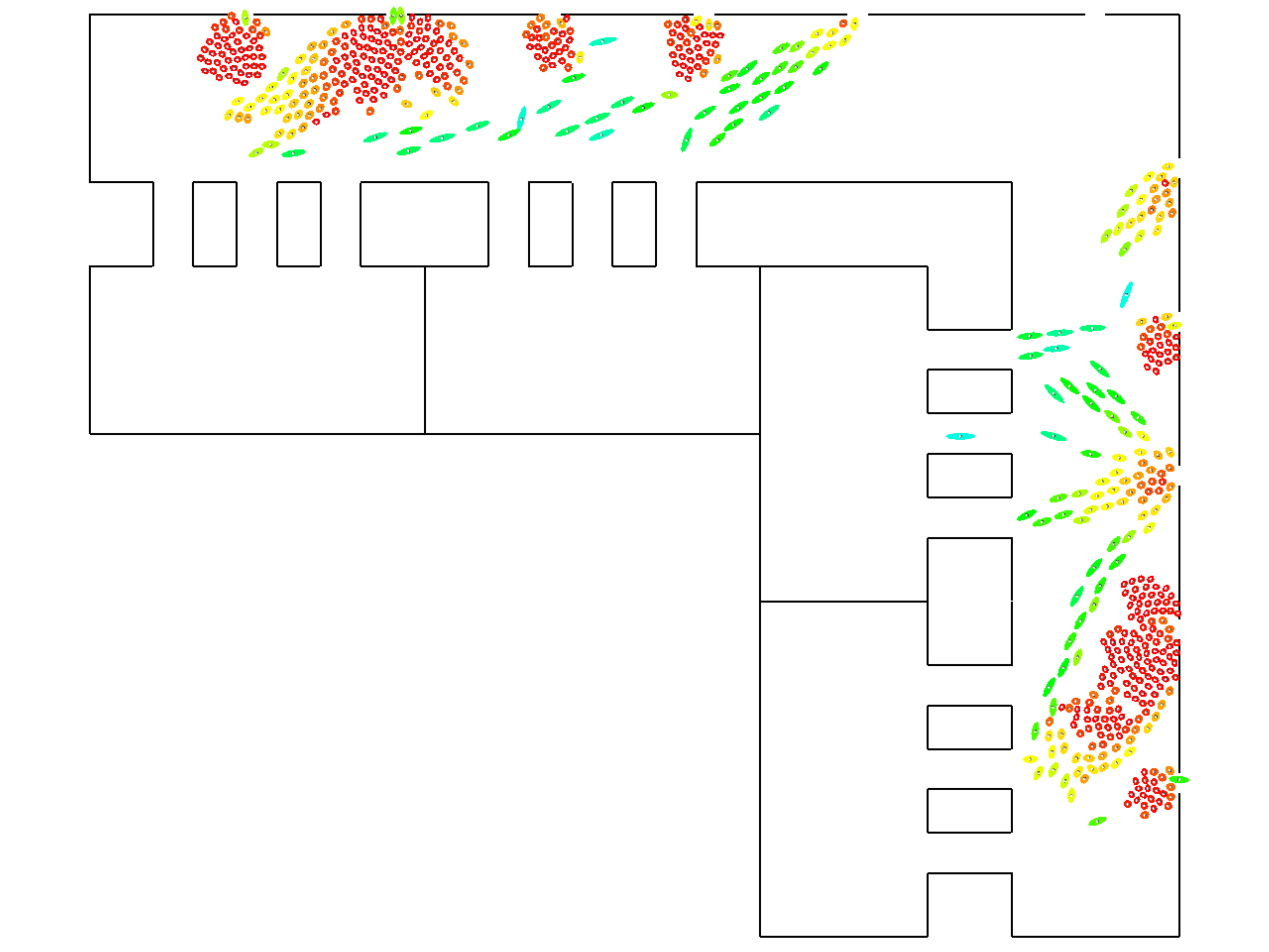}}
\caption[]{Dynamics of the system after 60 seconds for the initial distribution in Fig. \ref{fig:arena_complete}.  Congestions areas are red.}%
\label{fig:dynamic_complete}%
\end{figure}

\begin{figure}[htb]
\centering
\subfloat[Evacuation time distribution]{\label{fig:evac_time_complete}\includegraphics[width=50mm]{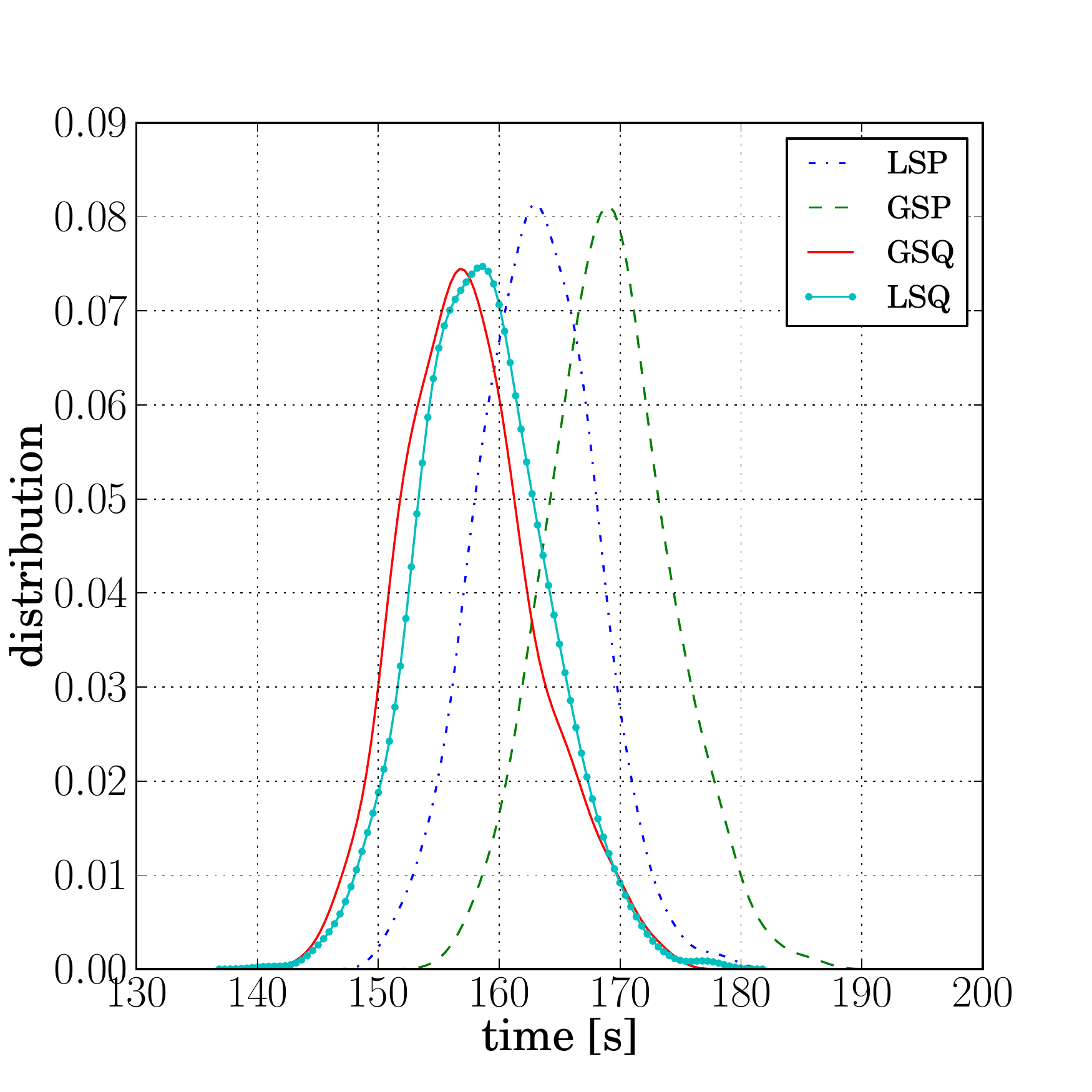}}
\subfloat[Jam time distribution]{\label{fig:jam_time_complete}\includegraphics[width=50mm]{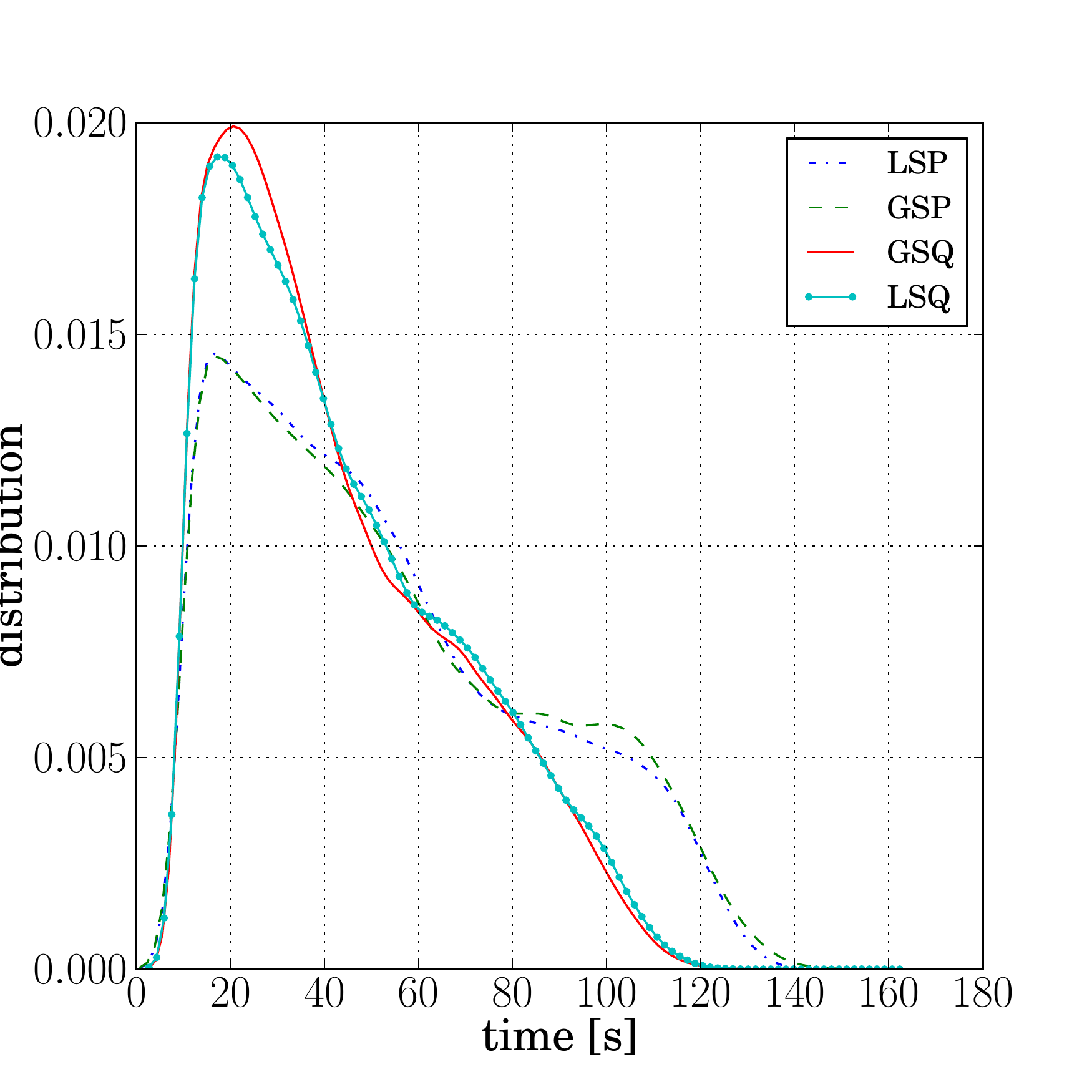}}
\subfloat[Jam size distribution]{\label{fig:1000_complete_jam_distribution}\includegraphics[width=50mm]{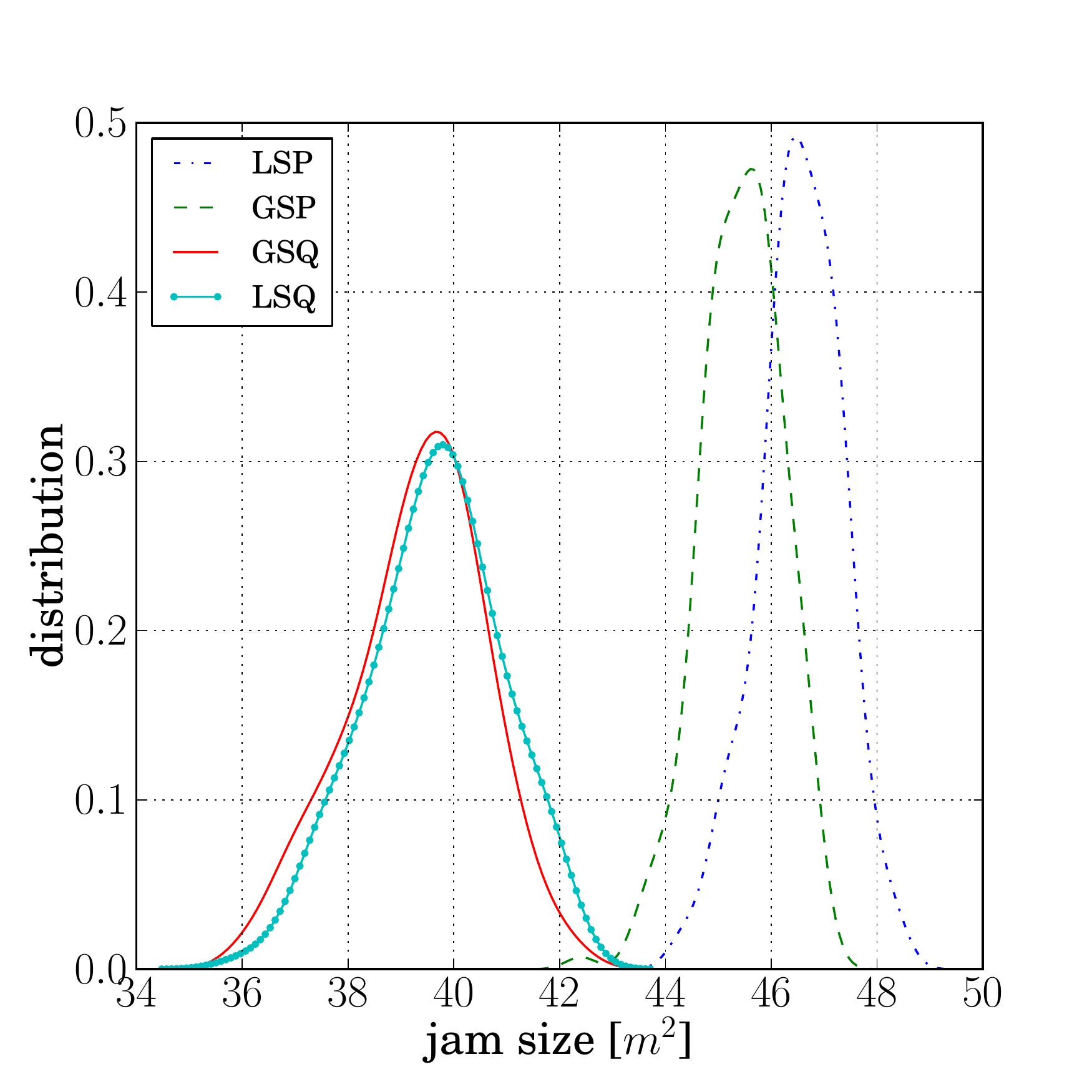}}
\caption{Evacuation time, time in jam  and jam size distribution for 1000 pedestrians. The initial positions are presented in Fig. \ref{fig:arena_complete}}%
\label{fig:results_complete}%
\end{figure}

The second test consists of the simulation of the evacuation of the complete facility with 1000 pedestrians. 
The initial distribution is shown in Fig. \ref{fig:arena_complete}. A snapshot of the simulation after 60 seconds 
is shown in Fig. \ref{fig:dynamic_complete}. The results of   the jamming time and the evacuation time are summarized 
in Fig. \ref{fig:evac_time_complete} and  Fig. \ref{fig:jam_time_complete}  respectively. There is no much difference between 
the mean values of the distributions as in the previous case. 
This is due to the highly congested situation. There is no such ''quickest path'' in this case and most of the pedestrians just have to follow the flow. 
Those results are confirmed by the time in jam distribution. The total jam size distribution is taken from Fig. \ref{fig:1000_complete_jam_distribution}. 
The mean jam sizes are 46 $m^2$ for the static and  40 $m^2$ for the dynamic strategies. 

\begin{figure}[htb]
\centering
\subfloat[Evacuation time distribution]{\label{fig:evac_time_disturbance}\includegraphics[width=50mm]{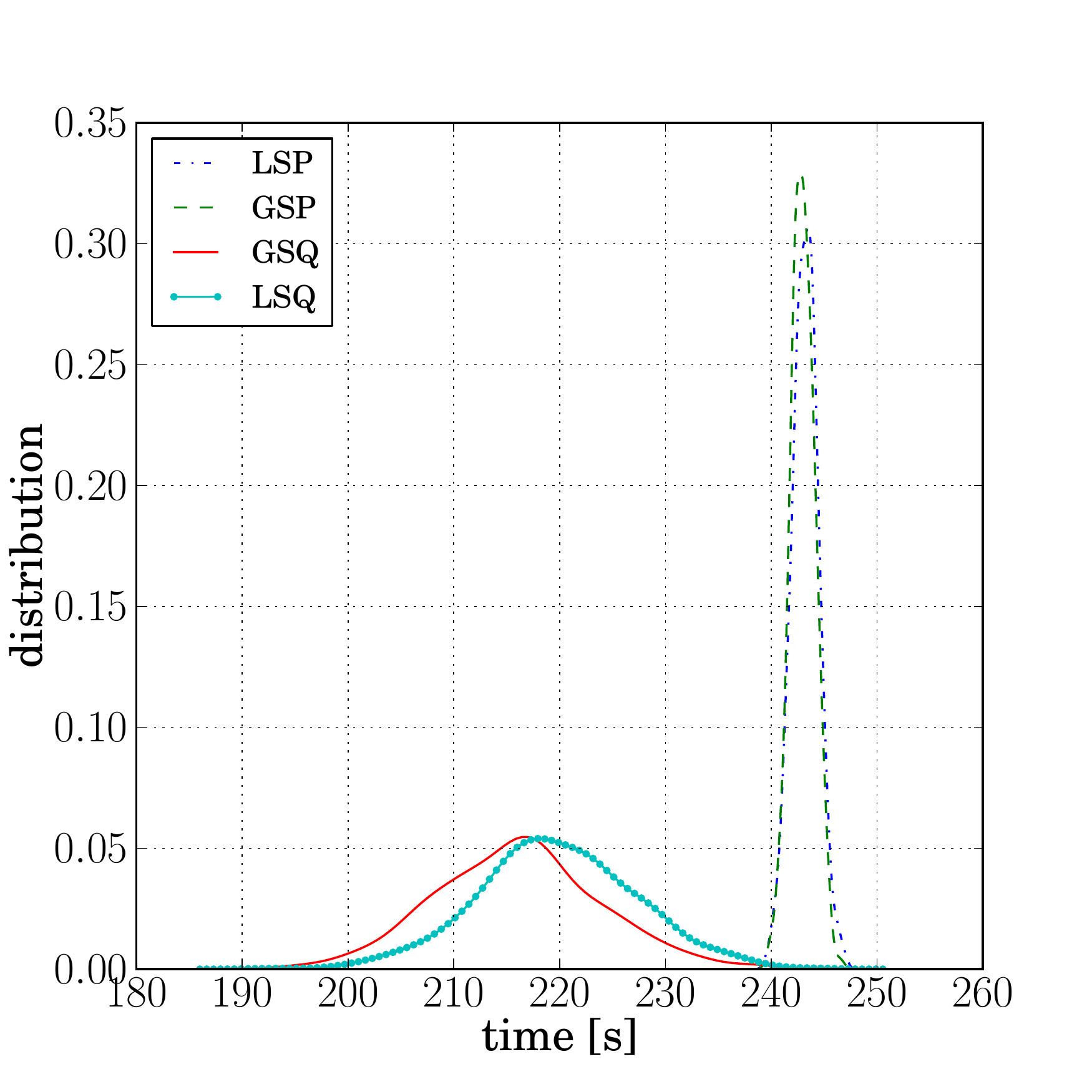}}
\subfloat[Jam time distribution]{\label{fig:jam_time_disturbance}\includegraphics[width=60mm, height=50mm]{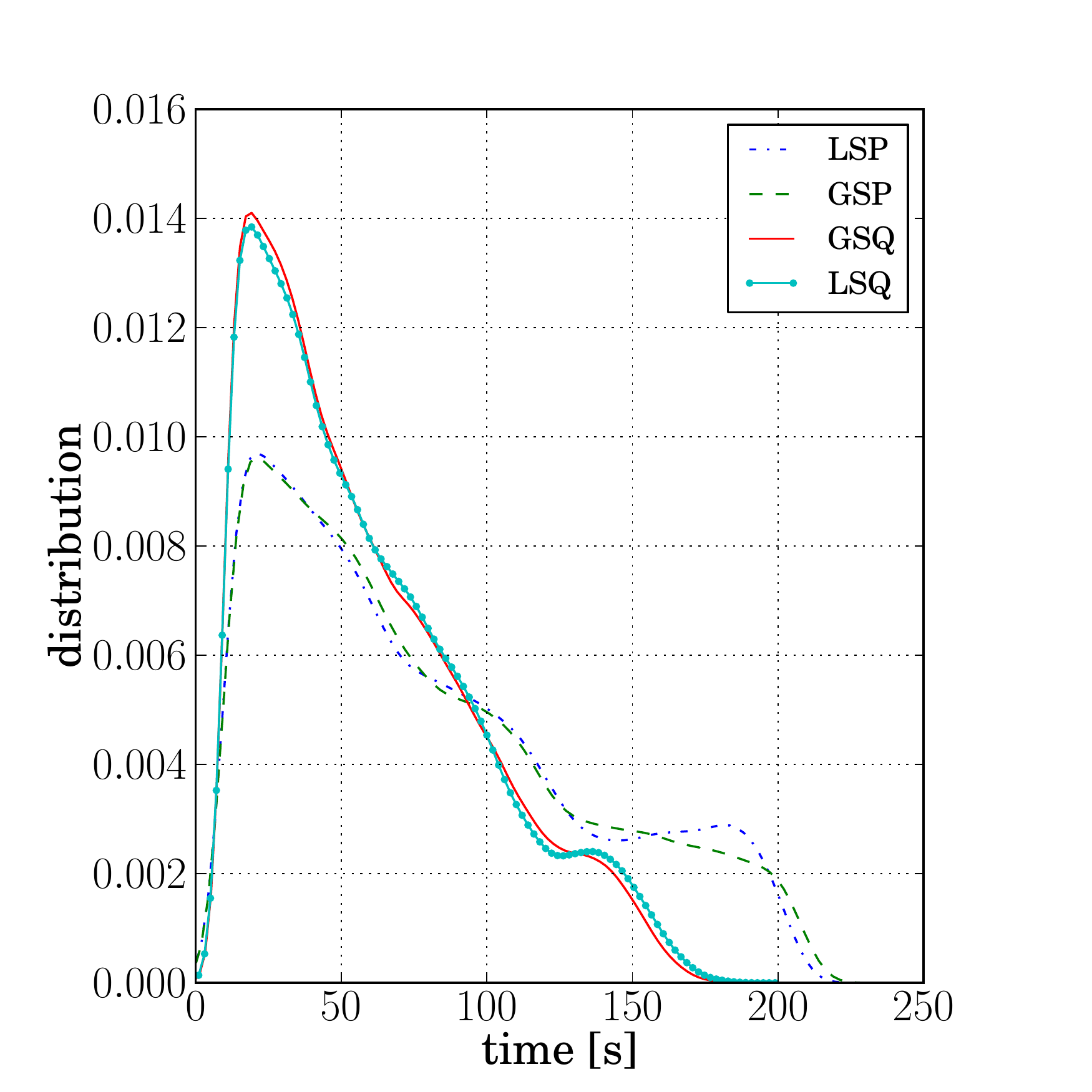}}
\subfloat[Jam size distribution]{\label{fig:1000_complete_jam_distribution_dist}\includegraphics[width=50mm]{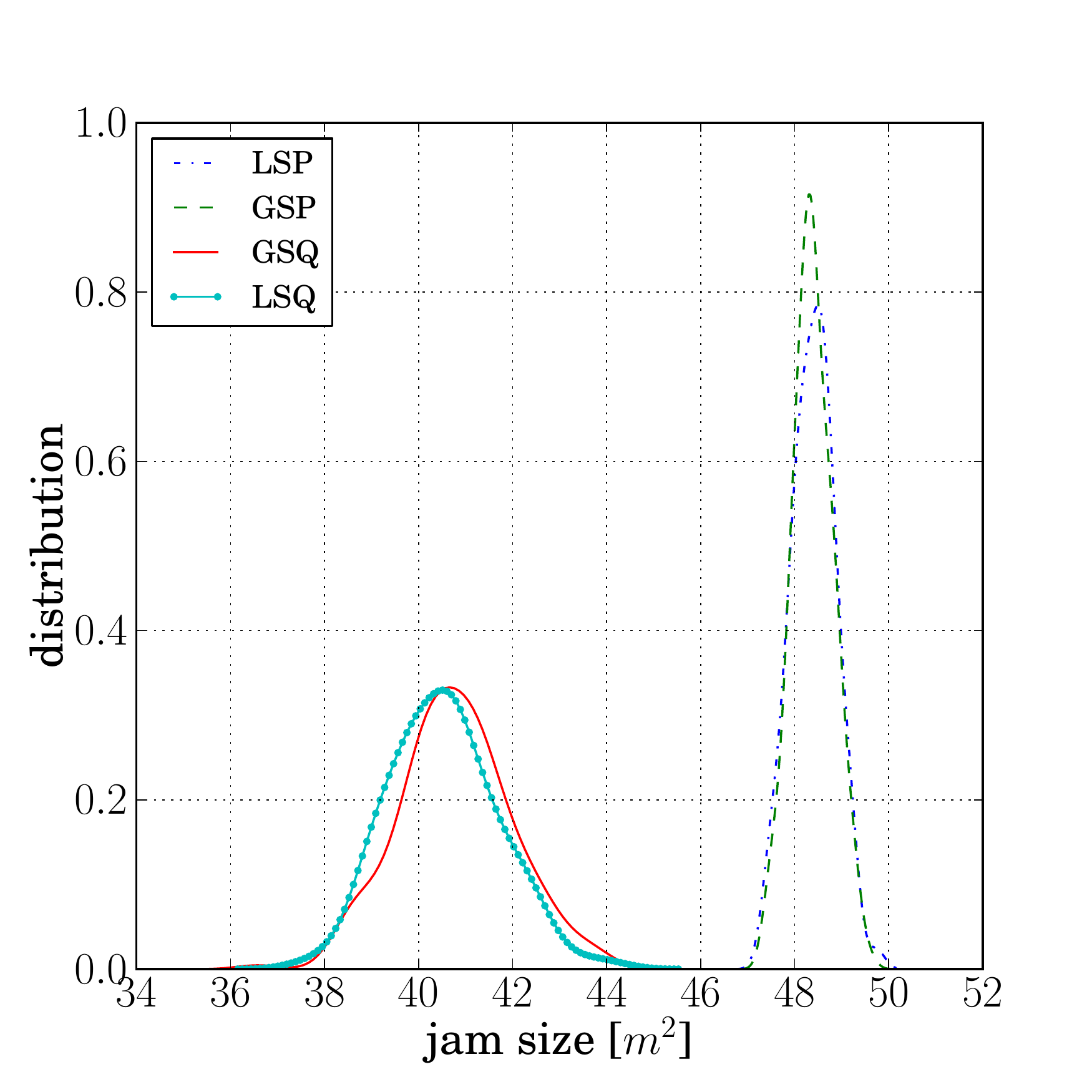}}
\caption{Evacuation time and time in jam distribution for 1000 pedestrians after a system disturbance. Three exits have been closed (broken escape route). 
The initial positions are presented in Fig. \ref{fig:arena_complete}}
\label{fig:results_disturbance}%
\end{figure}

In the third simulation scenario, the response of the system to a disturbance is simulated. The exits E2 and E8  (see Fig. \ref{fig:arena_single}) 
are broken and cannot longer be used (see Fig. \ref{fig:arena_complete}). The results of evacuation time and time in jam are presented 
in Fig. \ref{fig:evac_time_disturbance} and Fig. \ref{fig:jam_time_disturbance} respectively. There is now a certain asymmetry 
in the escape route scheme and this is reproduced by the modelling approach. The quickest path  lead to a faster evacuation and 
to less time in jam. This behaviour is also the expected one. The total jam size distribution is taken from Fig. \ref{fig:1000_complete_jam_distribution_dist}. 
All mean values are lower than the previous case without disturbance.
\clearpage

\section{Summary}

The approach presented in this paper offers the possibility to assess a given building structure 
 considering the manifold of possible route choices. 
Four strategies have been presented in a combination of quickest and shortest path.
The quickest path approach, which is based on an observation principle, 
is not sensitive to initial distribution of pedestrians or special topologies like symmetric exits, making it quite general. 
Furthermore it leads to a more realistic dynamics in the evacuation simulation. In addition criteria to assess 
the criticality of an evacuation simulation have been elaborated.
We investigated the evacuation time distribution, the time in jam distribution and average jam size of
 pedestrians using a quickest and shortest path routing approaches in a graph-based way finding algorithm. 
 The approaches have been tested on different scenarios with different complexities involving symmetric, 
asymmetric or even broken  escape routes. Similarities between the distributions, 
their meanings and impacts on different evacuation scenarios have been quantitatively analysed.

\section*{Acknowledgements}
This work has been performed within the program “Research for Civil Security” in the field “Protecting and Saving Human Life" 
funded by the German Government, Federal Ministry of Education and Research (BMBF). The project is granted under the Grant-Nr.: 13N9952.
\vspace{1cm}

\bibliographystyle{elsarticle-num}

\end{document}